\documentclass[12pt]{iopart}

\usepackage[utf8]{inputenc} 

\usepackage{iopams}

\usepackage{units}

\usepackage{notes2bib}

\usepackage{graphicx} \newcommand{\ket}[1]{\left|#1\right>}
\newcommand{\var}{\mathrm{var}} \newcommand{\cov}{\mathrm{cov}}
\newcommand{\text}[1]{\textrm{\footnotesize #1}}

\usepackage{hyperref}
\usepackage{srcltx}
\begin{document}

\title{Entanglement-assisted atomic clock beyond the projection noise
  limit}
\author{Anne Louchet-Chauvet, J\"urgen Appel, Jelmer J.
  Renema, Daniel Oblak, Niels Kj\ae rgaard\footnote{Present address: Danish Fundamental Metrology, Matematiktorvet 307, 2800 Kgs-Lyngby, Denmark}, Eugene S. Polzik}
\address{QUANTOP, Niels Bohr Institute, University of Copenhagen,
Blegdamsvej 17, 2100 København Ø, Denmark}
\ead{polzik@nbi.dk}

\begin{abstract}
  We use a quantum non-demolition measurement to generate a spin
  squeezed state and to create entanglement in a cloud of $10^5$ cold
  cesium atoms, and for the first time operate an atomic clock
  improved by spin squeezing beyond the projection noise limit in a
  proof-of-principle experiment.  For a clock-interrogation time of
  $\unit[10]{\mu s}$ the experiments show an improvement of
  $\unit[1.1]{dB}$ in the signal-to-noise ratio, compared to the
  atomic projection noise limit.

\end{abstract}

\section{Introduction}
Atomic projection noise, originating from the Heisenberg uncertainty
principle, is a fundamental limit to the precision of spectroscopic
measurements, when dealing with ensembles of independent atoms. This
limit has been approached, for example in atomic
clocks~\cite{santarelli1999,wilpers2002,Ludlow:2008_SrLatticeClock}.
Theoretical studies have shown that introducing quantum correlations
between the atoms can help overcome this limit and reach even better
precision~\cite{wineland1992,huelga1997,giovannetti04:_quant_enhan_measur,andre2004,meiser2008}.
Spin squeezing in a system of two ions has been shown to improve the
precision of Ramsey spectroscopy for frequency measurements
\cite{meyer01:_ion_spin_squeezing}.  Furthermore, squeezed atomic
ensembles improve the sensitivity of
magnetometers~\cite{wasilewski:_magnetometry,koschorreck-2009:_magnetometry}.

In a previous publication~\cite{appel2009}, we have reported the
generation of quantum noise squeezing on the cesium clock transition
via quantum non-demolition (QND) measurements. By proposing an
entanglement-assisted Ramsey (EAR) method including a QND
measurement, we showed how this squeezing could help improve the
precision of atomic clocks. In this work, we describe the spin
squeezing experiment in more detail and implement the complete
EAR clock sequence. For the first time we demonstrate an
atomic microwave clock improved by spin squeezing.  Decoherence
effects are measured and included in the analysis. The clock reported
here does not reach record precision due to technical reasons, however
the demonstrated approach is applicable to the state-of-the-art
clocks, as indicated in~\cite{lodewyck:Nondestructivemeasurement}.

\section{Generation of a conditionally squeezed atomic state}

\subsection{Coherent and squeezed spin states}

An ensemble of $N_A$ identical 2-level atoms can be described as an
ensemble of pseudo-spin-$1/2$ particles. We define the collective
pseudo-spin vector $\hat{J}$ as the sum of all individual spins.
Traditionally, its $z$-component $J_z$ is defined by the population
difference $\Delta N$, such that: $J_z=\frac 1 2
(N_\uparrow-N_\downarrow)=\Delta N/2$.  A coherent spin state (CSS) is
a product state (i.e. atoms are uncorrelated) where the spins of $N_A$ atoms are aligned in the same
direction, for example such that $J_x=N_A/2$, i.e. $\ket{\mathrm{CSS}} =
\bigotimes_{i=1}^{N_A} \frac{1}{\sqrt{2}}
\left(\ket{\uparrow}_i+\ket{\downarrow}_i\right)$. Then, the other
projections of $\hat{J}$ minimize the Heisenberg uncertainty relation:
$\var(J_z)\cdot \var(J_y) \geq \left<J_x\right>^2/4$ and
$\var(J_z)=\var(J_y)=N_A/4$. These quantum fluctuations, referred to as
\emph{CSS projection noise}, pose a fundamental limit to the precision
of the $J_z$ measurement~\cite{itano1993}.  It is possible to reduce
the fluctuations of one of the spin components - for example $J_z$ -
to below the projection noise limit by introducing quantum
correlations between different atoms within the atomic ensemble. In
this case, the fluctuations on the conjugate observable - here $J_y$ -
increase according to the Heisenberg uncertainty relation. Such a
state is referred to as a spin squeezed state (SSS). Whether the
atoms exhibit non-classical correlations is determined by the criterion
\begin{equation}
  \var(J_z)<\frac{\left<J\right>^2}{N_A}
  \qquad
  \Leftrightarrow
  \qquad
  \xi = \frac{\var(J_z)}{\left<J\right>^2}N_A  < 1
  \label{eq:wineland},
\end{equation}
where $\xi$ is called the squeezing parameter. Under this condition
(even for a general mixed state) the atoms are entangled whereby the signal-to-projection-noise ratio in spectroscopy and
metrology experiments is improved by a factor of $1/\xi$ in variance, or $1/\sqrt{\xi}$ in standard deviation~\cite{wineland1992}.
Equation~(\ref{eq:wineland}) will be referred to as the Wineland
criterion throughout this paper.

Spin squeezing can be produced, for example, by atomic
interactions~\cite{sorensenduan2001:bec_entanglement,esteve2008:bec_spinsqueezing},
by mapping the properties of squeezed light onto an atomic
ensemble~\cite{kuzmich97:_spin_squeez_ensem_atoms_illum_squeez_light,hald99:_spin_squeez_by_light,
  appel2008:quantum_memory_squeezed,honda:storage_squeezed}, or by
non-destructive measurements on the atoms~\cite{grangier91:QND,
  kuzmich98:_atomicQND,kuzmich00:_spinsqueez_continuous,
  chaudhury06:_contin_nondem_measur_cs_clock_trans_pseud,SchleierSmith2008,
  takano09:_spin_squeez_cold_atomic_ensem,julsgaard01:_entanglement}.
We follow the latter approach by performing a weak, non-destructive
measurement of the $J_z$ spin component.  Any later measurement on
$J_z$ on the same ensemble will be partly correlated to the first
measurement outcome.  Therefore, the outcome of a subsequent
$J_z$-measurement can be predicted to a precision better than the
CSS-projection noise.  In other words, if $\phi_1$ and $\phi_2$ are
the outcomes of the first and second measurements, respectively, the
conditional variance $\var(\phi_2-\zeta \phi_1)$ is reduced to below
the variance of a single measurement $\var(\phi_1)=\var(\phi_2)$,
where $\zeta$ is the correlation strength
$\zeta\equiv\cov(\phi_1,\phi_2)/\var(\phi_1)$. If the QND measurement
does not reduce the length of the pseudo-spin vector $\langle J
\rangle$ too much, Eq.~(\ref{eq:wineland}) implies that the reduction
of the variance results in a metrologically relevant SSS.

\subsection{Preparation of the coherent spin state} \label{atomsprep}

\begin{figure} \centering
  \includegraphics[width=9cm]{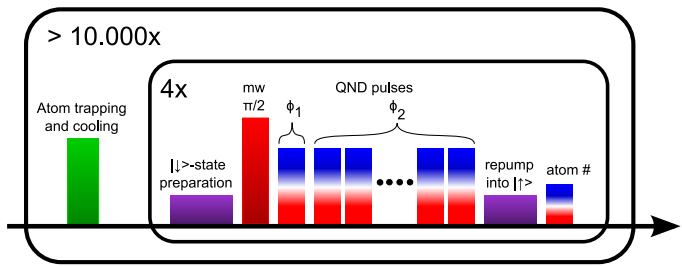}
  \caption{Experimental pulse sequence for the preparation of a coherent spin state and quantum non-demolition measurements.}
  \label{fig:seqsqueezing}
\end{figure}

The experimental sequence for the preparation of the coherent spin state
and the QND measurements is shown in Fig.~\ref{fig:seqsqueezing}.
Cesium atoms are first loaded from a background cesium vapor into a
standard magneto-optical trap (MOT) on the $D_2$-line, and are then
transferred into an elongated far off-resonant trap (FORT). The FORT is generated
by a Versadisk laser with a wavelength of \unit[1032]{nm} and a power
of $\unit[2.3]{W}$, which is focused to a $\unit[20]{\mu m}$ radius spot to
confine an elongated atomic sample.
After the loading of the FORT,
the MOT is switched off and a bias magnetic field is applied, defining
a quantization axis orthogonal to the trapping beam. The
$6S_{1/2}\ket{F=3, m_F=0}$ and $6S_{1/2}\ket{F=4, m_F=0}$ ground
levels are referred to as the \emph{clock levels}. We denote them as
$\ket{\downarrow}$ and $\ket{\uparrow}$, respectively. The cesium
atoms are then prepared in the clock level $\ket{\downarrow}$ by optical
pumping. Atoms remaining in states other than $\ket{\downarrow}$ due
to imperfect optical pumping are subsequently pushed out of the
trap, as described in~\cite{appel2009}. A resonant microwave pulse ($\pi/2$-pulse) is
used to put the atoms into $\ket{\mathrm{CSS}}$. We then perform successive QND
measurements of the atomic population difference $\Delta N$ by detecting the
state-dependent phase shift of probe light pulses with a Mach-Zehnder
interferometer as described in detail in section 2.3. Later on, we
optically pump the atoms into $F=4$ to measure the atom number
$N_A$. We recycle the remaining atoms for three subsequent
experiments, preparing them into a CSS, performing successive QND
measurements and finally measuring the atom number. After these four
experiments, all the atoms are blown away with laser light and we perform three series
of QND measurements with the empty interferometer to obtain a zero
phase shift reference measurement. This sequence is repeated several
thousand times with a cycle time of $\approx \unit[5]{s}$.

To ensure that the microwave does not adress hyperfine transitions
other than the clock transition, the bias magnetic field $B$ is set so
that the Zeeman splitting $\Delta E_Z$ between adjacent magnetic
sublevels ($\unit[350]{kHz/G}$ to first order) coincides with one of the
zeroes of the microwave pulses' frequency spectrum: $\Delta
E_Z=\ell/\tau_{\pi/2}$, where $\tau_{\pi/2}$ is the microwave $\pi/2$
pulse duration, and $\ell$ is an integer number. Therefore for typical
durations of $\tau_{\pi/2}=\unit[7]{\mu s}$, we set the bias magnetic field to
$\unit[1.22]{G}$.

\subsection{Dispersive population measurement}
\begin{figure} \centering
  \includegraphics[width=8cm]{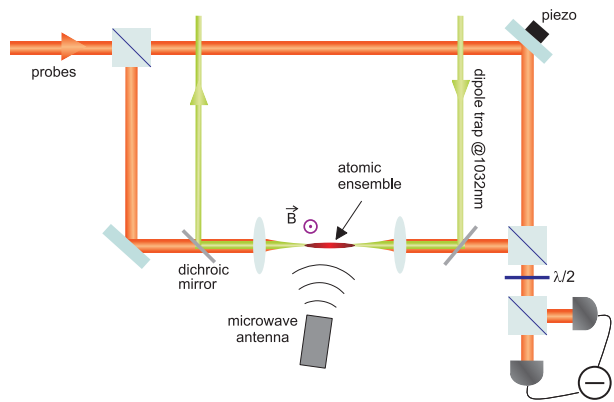}
  \caption{Mach-Zehnder interferometer. The atoms are placed in one
    arm of a Mach-Zehnder interferometer. The dipole trapping beam is
    overlapped with the probe arm to form an elongated atomic cloud
    along the propagation direction of the probe beam. Two probe beams
    of identical linear polarization enter the interferometer via the
    same input port and acquire phase shifts proportional to the
    number of atoms in the clock states $N_\uparrow$ and
    $N_\downarrow$, respectively. The bias magnetic field (B) is
    aligned to the polarization of the probe beams.}
  \label{fig:mzi}
\end{figure}

\begin{figure} \centering
  \includegraphics[width=9cm]{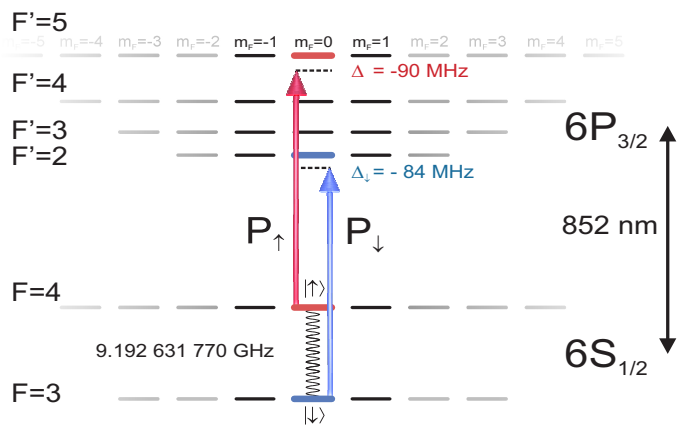}
  \caption{Cesium $D_2$-line level diagram}
  \label{fig:levels}
\end{figure}

In order to measure an atomic squeezed state, we require a measurement
sensitivity that is sufficient to reveal the atomic projection noise
limit.  The population in each clock level is measured via the phase
shift imprinted on a dual-color beam propagating through the atomic
cloud~\cite{saffman09:_spinsqueez_multicolor}.  The dipole trap is
overlapped with one arm of a Mach-Zehnder interferometer (see
Fig.~\ref{fig:mzi}). A beam $P_\downarrow$ of one color off-resonantly
probes the $\ket{F=3}\rightarrow\ket{F'=2}$ transition, whereas a
second beam $P_\uparrow$ off-resonantly probes the
$\ket{F=4}\rightarrow\ket{F'=5}$ transition (see
Fig.~\ref{fig:levels}). Each color experiences a phase shift
proportional to the number of atoms in the ground state of the probed
transition: $\phi_\downarrow=\chi_\downarrow N_\downarrow$ and
$\phi_\uparrow=\chi_\uparrow
N_\uparrow$~\cite{windpassinger08:_nondes_probin_rabi_oscil_cesium}.
We carefully choose the probe detunings to ensure that the coupling
constants are equal: $\chi_\uparrow=\chi_\downarrow$. The two
probe beams emerge from one single-mode polarization maintaining
fiber, and their intensities are stabilized to be equal
$n_\uparrow=n_\downarrow$ to within $0.1\%$.

For each individual probe, the photocurrent difference between the
detectors at the two output ports of the interferometer reads as:
\begin{equation}
  \Delta n_\uparrow=n_\uparrow \beta \cos(k_\uparrow \Delta L+\chi_\uparrow
  N_\uparrow) \textrm{ and } \Delta n_\downarrow=n_\downarrow \beta \cos(k_\downarrow
  \Delta L+\chi_\downarrow N_\downarrow),
\end{equation}
where $k_\uparrow,k_\downarrow$ are the wave vector lengths for
$P_\uparrow,P_\downarrow$, respectively, $\Delta L$ is the
interferometer path length difference and $\beta=\sqrt{13}$ is the ratio of the
field amplitudes in the reference- and probe-arm of the interferometer.
In the absence of atoms $N_\uparrow=N_\downarrow=0$; therefore the
smallest interferometer path length difference which leads to opposite
phase for the signals $\Delta n_\uparrow$ and $\Delta n_\downarrow$
is: $\Delta L_0 = \frac{\pi}{k_\uparrow - k_\downarrow}$.  Since the
wavelengths of the two probe lasers are so close (to one part in
$40\,000$), one can assume that the fringes of each color are of
opposite phase in the neighborhood of several wavelengths around
$\Delta L_0$.  The interferometer path length
difference 
\bibnote{In \cite{appel2009} we used opposite input ports for the two probe colors and operated the interferometer in its white-light position $\Delta L=0$ to minimize sensitivity to differential probe frequency noise. The method described here allows us to feed light of both probe colors into the interferometer through one common single mode fiber. This eliminates a possible spatial mode mismatch, which leads to spatially inhomogeneous differential AC-Stark shifts across the atomic sample. The differential probe frequency is controlled tightly using an optical phase lock \cite{appel09:_pll}.} 
is set to the value closest to
$\Delta L_0$ that also satisfies: $\Delta n_\uparrow=\Delta
n_\downarrow=0$. This path length difference therefore varies with the
expected atom number. To account for technical imperfections in the
balancing of the probe powers and frequencies we define the average
effective coupling strength $n \chi \equiv (n_\uparrow \chi_\uparrow +
n_\downarrow \chi_\downarrow) /2 $ and the balancing error $n \Delta
\chi \equiv (n_\uparrow \chi_\uparrow - n_\downarrow \chi_\downarrow)
/2$ as well as the average intensity $n=(n_\uparrow+n_\downarrow)/2$.

This way, the total photocurrent difference
$\Delta n=\Delta n_\uparrow+ \Delta n_\downarrow$ reads as:
\begin{equation} \Delta n=n_\uparrow \beta \sin (\chi_\uparrow N_\uparrow) -n_\downarrow \beta \sin (\chi_\downarrow
  N_\downarrow) \simeq n \beta (\chi\, \Delta N + \Delta \chi \, N_A).
\end{equation}

We define the phase measurement outcome as:
\begin{equation} \phi \equiv \frac{\Delta n}{\beta n} =\frac{\delta n}{\beta n} + \chi\,
  \Delta N + \Delta \chi \, N_A,  
\end{equation}
where $\delta n$ denotes the total shot noise contribution from both
colors. The phase $\phi$ provides a measurement of $\Delta N$ with
added shot noise and classical noise. For $N_A$ atoms in a CSS, we
have $\var(\Delta N)=N_A$.  The projection noise increases with the
atom number and using $\langle \Delta \chi \rangle =0, \langle \Delta
N \rangle =0$ we obtain:
\begin{equation} \var(\phi)=\frac{\beta^2 +1}{2 \beta^2} \, \frac{1}{n} + \chi^2 N_A + \var(\Delta \chi) \, {N_A}^2.
  \label{eq:varphi}
\end{equation}

After each experiment we use the same dual-color probe beam to
determine the total atom number $N_A$. To that end we first optically
pump all atoms into $\ket{F=4,m_F=-1,0,+1}$. The phase
measurement outcome reads as $\phi=n\bar{\chi} N_A$, where
$\bar{\chi}$ is the effective coupling constant for the probe
$P_{\uparrow}$, when the atoms are distributed among different
magnetic sublevels. The similarity of the Clebsch-Gordan coefficients
for the $\ket{F=4,m_F}\rightarrow\ket{F'=5,m_F}$ transitions (with low
$|m_F|$) ensures that $\bar{\chi}\approx \chi$ which is confirmed by
experiments with a precision of better than $\unit[5]{\%}$.


\subsection{Projection noise measurements}

The two probe beams are generated by two extended-cavity diode lasers, which are phase-locked in order to minimize their relative frequency
noise~\cite{appel09:_pll}. Their detunings are given in
Fig.~\ref{fig:levels}.

\begin{figure} \centering
  \includegraphics[width=7cm, angle=-90]{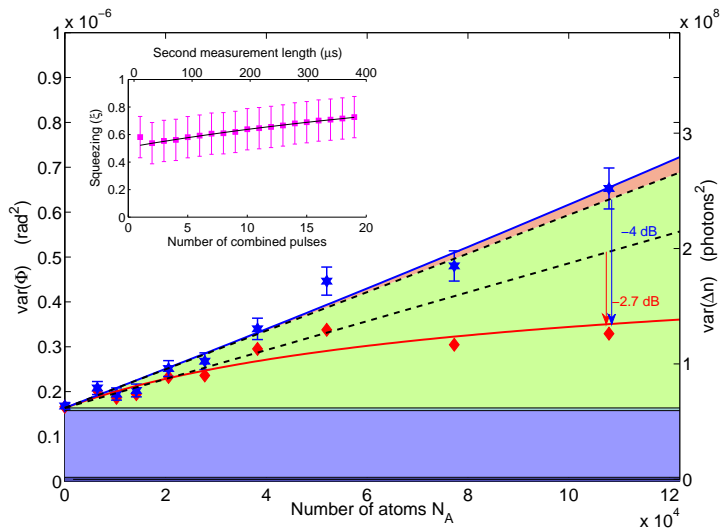}
  \caption{Projection noise and spin squeezing. The first QND
    measurement is performed with a $10~\mu$s bichromatic pulse
    containing $6\cdot10^6$ photons in total. The second measurement
    is comprised of two such pulses. Blue stars: variance of the
    second measurement $\var(\phi_2)$, where the data are sorted by
    atom number and grouped into 10 bins. Solid blue line: quadratic
    fit to $\var(\phi_2)$. Green area: atomic projection noise of the
    CSS. Red diamonds: conditional variance $\var(\phi_2-\zeta
    \phi_1)$. Red line: reduced noise as predicted by the fits to the
    noise data. The error bars correspond to the statistical
    uncertainty given by the number of measurements. This data is
    obtained by acquiring 1200 MOT loading cycles, i.e. 4800
    experiments. Inset: metrologically relevant spin squeezing $\xi$
    (as defined in Eq.~\ref{eq:wineland}) as a function of
    the number of pulses combined to form the second measurement. The
    data is fitted with an exponential decay (solid line).}
  \label{fig:squeezing}
\end{figure}

In Fig.~\ref{fig:squeezing}, we analyze the variance of the atomic population
difference measurement as a function of the atom number.

As the data acquisition proceeds over several hours, it is a major
challenge to keep experimental parameters such as $\tau_{\pi/2}$
constant to much better than $1/\sqrt{N_A}$ so that the mean values of
the atomic population difference measurements would not drift by more
than their quantum projection noise. To eliminate the influence of
such slow drifts, we subtract the outcomes of measurements
performed on independent atomic ensembles recorded in successive MOT
cycles from each other. We then calculate the variances using this
differential data.

The correlated and uncorrelated parts of the noise variance are fitted
with second order polynomials. According to Eq. \ref{eq:varphi}, we
interpret the linear part of this fit as the CSS projection noise
contribution. We observe a negligible quadratic part which means that
classical noise sources like laser intensity and frequency
fluctuations, which cause noise in the effective coupling constant are
small.  Achieving this linear noise scaling is a significant
experimental challenge (see \cite[Suppl.]{appel2009}), since the
effect on $\Delta N$ of various sources of classical noise must be
kept well below the level of $1/\sqrt{N_A}\simeq 3\cdot 10^{-3}$
between independent measurements, i.e. over a duration of
$\simeq\unit[5]{s}$.

\subsection{Conditional noise reduction}

Both measurements $\phi_1$ and $\phi_2$ are randomly normally
distributed around zero with variances that have contributions from the
shot noise and the atomic projection noise:
\begin{equation} \var(\phi_1)=\frac{1}{n_1} + \chi^2 N_A, \
  \var(\phi_2)=\frac{1}{n_2} + \chi^2 N_A.
\end{equation}
Since the two measurements are performed on the same atomic sample,
they are correlated: $\cov(\phi_1,\phi_2) = \chi^2 N_A$. The
conditional variance $\var(\phi_2-\zeta \phi_1)$ is minimal when
$\zeta =\frac{\cov(\phi_1,\phi_2)} {\var(\phi_1)}$:
\begin{equation} \var(\phi_2-\zeta \phi_1)=\frac{1}{n_2} +
  \frac{1}{1+\kappa^2} \chi^2 N_A.
\end{equation}

We measure $\kappa^2=1.6$ for $N_A=1.2\cdot10^5$ atoms and we obtain a
reduction of the projection noise by $\frac{1}{1+\kappa^2}=-4$~dB
compared to the CSS projection noise, as indicated by the blue arrow
in Fig.~\ref{fig:squeezing}.

\subsection{Decoherence}
\label{sec:decoherence}

Dispersive coupling is inevitably accompanied by spontaneous photon
scattering, inducing atomic population redistribution among the ground
magnetic sublevels, as well as a reduction of coherence between the
clock levels. Due to the selection rules, the population
redistribution predominantly occurs within the Zeeman structure of the
hyperfine levels~\cite{saffman09:_spinsqueez_multicolor}: Atoms that
scatter a photon from the $P_\downarrow$ probe almost certainly end up
in the $\ket{F=3,m_F=-1,0,+1}$ sublevels again and atoms that
scatter a photon from the $P_\uparrow$ probe predominantly end up in
the $\ket{F=4,m_F=-1,0,+1}$ states (see. Fig.~\ref{fig:levels}). More
importantly, the similarity of the Clebsch-Gordan coefficients that
describe coupling of the probe light to low $|m_F|$-sublevel states
ensures that the optical phase shift is almost unaffected by such a
population redistribution. This makes our dual-color measurement an
almost ideal QND-measurement as spontaneous scattering events
effectively do not change the outcome of a $J_z$ measurement, i.e. no
extra projection noise due to repartition into the opposite hyperfine ground
states is added.  On the other hand, the spontaneous photon scattering
still leads to a shortening of the mean collective spin vector $ \langle J
\rangle \rightarrow (1-\eta) \langle J \rangle $.

\begin{figure} \centering
\footnotesize{(a)}  \includegraphics[width=8cm]{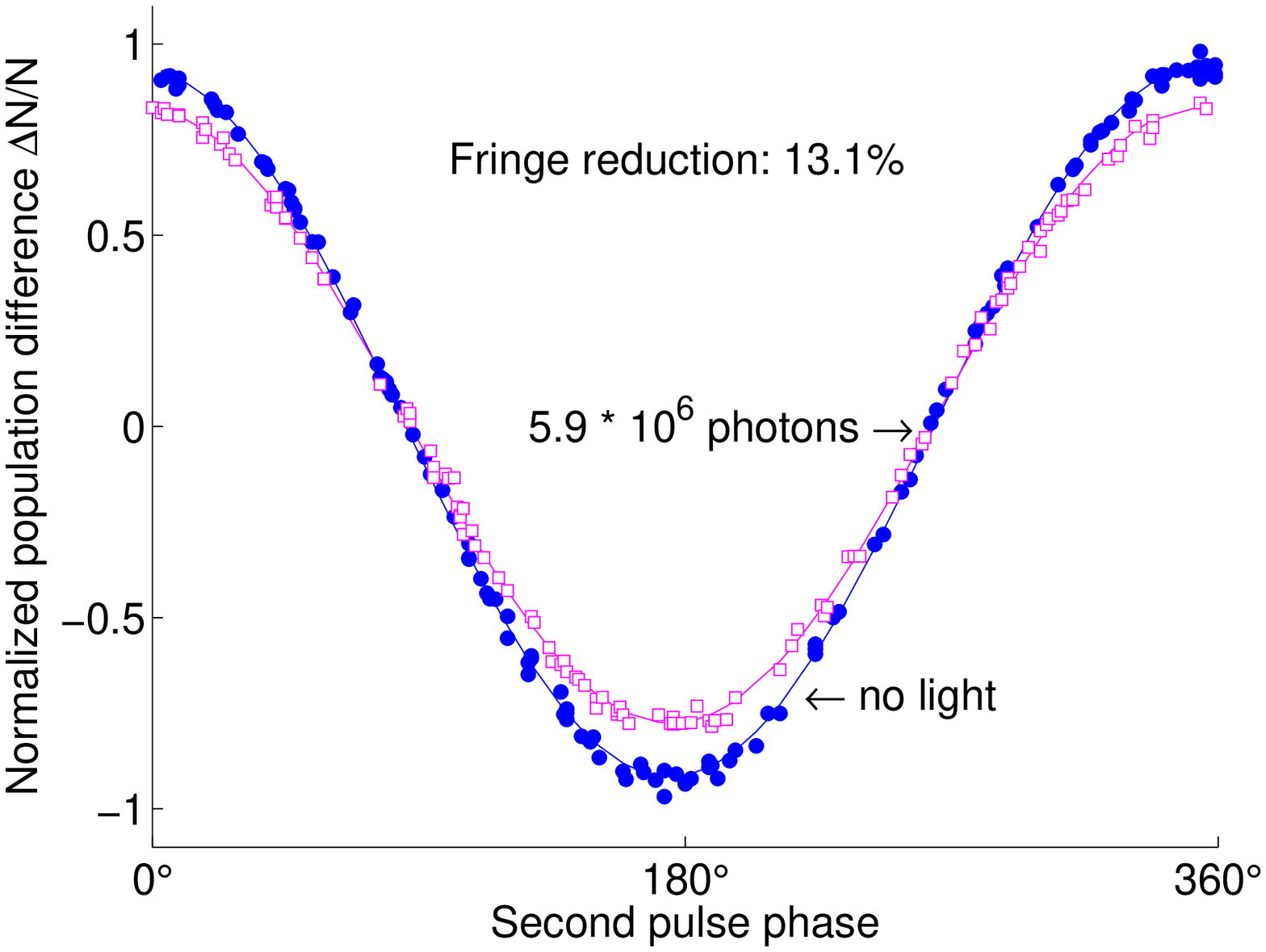}
  \includegraphics[width=6cm]{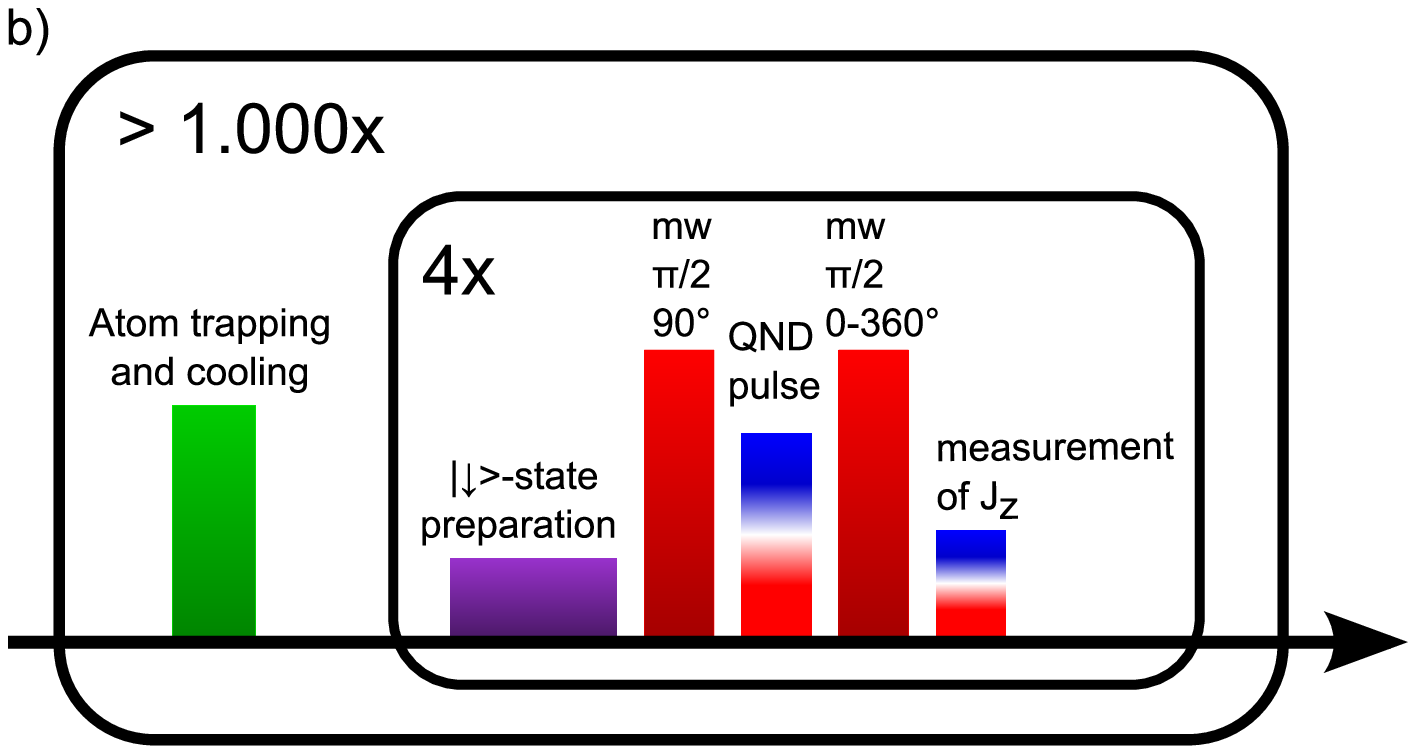}
  \caption{Decoherence measurement via Ramsey fringe reduction. (a)
    Blue points: full Ramsey fringe. Red points: reduced fringe when a
    QND-pulse containing $5.9\cdot 10^6$ photons is inserted in the
    Ramsey sequence. In this experiment, the fringe height is reduced
    by $13.1\%$, leading to a decoherence parameter
    $\alpha=2.39\times10^{-8}$. (b) Pulse sequence. }
  \label{fig:decoherence}
\end{figure}

The photon-number dependence of this mechanism can be modeled as
$\eta=1-e^{\alpha n}$, where $n$ is the total number of photons in the
probe pulse. We measure the parameter $\alpha$ in a separate
experiment, by comparing the Ramsey fringe amplitudes with and without
a bichromatic QND pulse between the two microwave $\pi/2$-pulses, as
shown in Fig.~\ref{fig:decoherence}.  We obtain a value of
$\alpha=-2.39\cdot 10^{-8}$ for this particular atomic cloud geometry.

Each probe color induces an inhomogeneous ac Stark shift on the atomic
levels, causing additional dephasing and decoherence.  Nevertheless,
in our two-color probing scheme, the Stark shift on level
$\ket{\uparrow}$ caused by probe $P_{\uparrow}$ is compensated by an
identical Stark shift on level $\ket{\downarrow}$ caused by probe
$P_{\downarrow}$, provided the probe frequencies are set so that
$\chi_\uparrow=\chi_\downarrow$, and the probe powers are equal.
Hence, the two Ramsey fringes depicted on Fig.~\ref{fig:decoherence}
(with and without a light pulse) are in phase to better than
$1^\circ$.

\subsection{Squeezing and entanglement}

A QND measurement is characterized by the decoherence $\eta$ it
induces, and by the $\kappa^2$-coefficient that describes the
measurement strength. In the absence of classical noise, for the
dual-color QND the squeezing parameter as defined in
Eq.~\ref{eq:wineland} can be written as:

\begin{equation} \xi_\text{lin}=\frac{1}{(1-\eta)^2\, (1+\kappa^2)}.
  \label{eq:etaeqn}
\end{equation}

Therefore, for a fixed detuning of the laser frequencies, the choice of the number of photons used in a QND
measurement is the result of a trade-off between the amount of
decoherence induced by the photons and the amount of information that
the $\phi_1$-measurement yields~\cite{windpassinger09:_squeez}.

Here, we use $n_1/2=3\cdot 10^6$ photons per probe color, inducing a
shortening of $\eta = 14\%$ of the collective spin $J$.  From Eq.
\ref{eq:etaeqn} we expect an improvement of the
signal-to-projection-noise ratio by $1/\xi_\text{lin}=2.8$~dB compared
to a CSS. From the data presented in Fig.~\ref{fig:squeezing} we
measure $1/\xi=2.7$~dB, as indicated with a red arrow. This value
agrees well with the theory and with the results reported
in~\cite{appel2009}.

The detuning of the probe light with respect to the atomic transitions
(see Fig.~\ref{fig:levels}) as well as the duration of the probe pulses
($\unit[10]{\mu s}$) have been chosen to minimize the influence of
classical noise sources such as electronic noise in the detector and
relative frequency- and intensity-noise of the two probe colors.

Since the probe pulses redistribute population only within the
hyperfine manifold (cf. sec. \ref{sec:decoherence}), the light shot
noise contribution in the second measurement variance can be reduced
by increasing the number of photons $n_2$ in the second measurement
(cf. Eq. \ref{eq:etaeqn}). In our experiment, the second measurement
is comprised of $\unit[10]{\mu s}$ long bichromatic pulses containing
a total of $n_2$ photons, separated by $10~\mu$s.  As shown in the
inset of Fig.~\ref{fig:squeezing}, $\xi$ exponentially approaches
unity as more pulses are combined to form the second measurement.  We
attribute this decay to the atomic motion within the dipole trap
during the second measurement: the atoms move in the transverse
profile of the probe beam, and are probed with a position dependent
weight corresponding to the local probe light intensity. Movement during the time 
 interval between the first and second probe pulse therefore induces a decay of
the correlations between the two measurements.  We fit
the experimental data with:
\begin{equation} \xi(t_2)=1-B e^{-t_2/\tau_\mathrm{decay}},
\end{equation}
where $t_2$ is the total duration of the second measurement, and
obtain $\tau_\mathrm{decay}=\unit[670]{\mu s}$, which is of the same order of
magnitude as half the radial trap oscillation
period~\cite{oblak08:_echo_gauss}.

\section{Entanglement-assisted atomic clock}

\subsection{Entanglement-assisted Ramsey sequence}

\begin{figure}[ht] \centering
  \includegraphics[bb=49 116 790 326,clip,width=14.5cm]{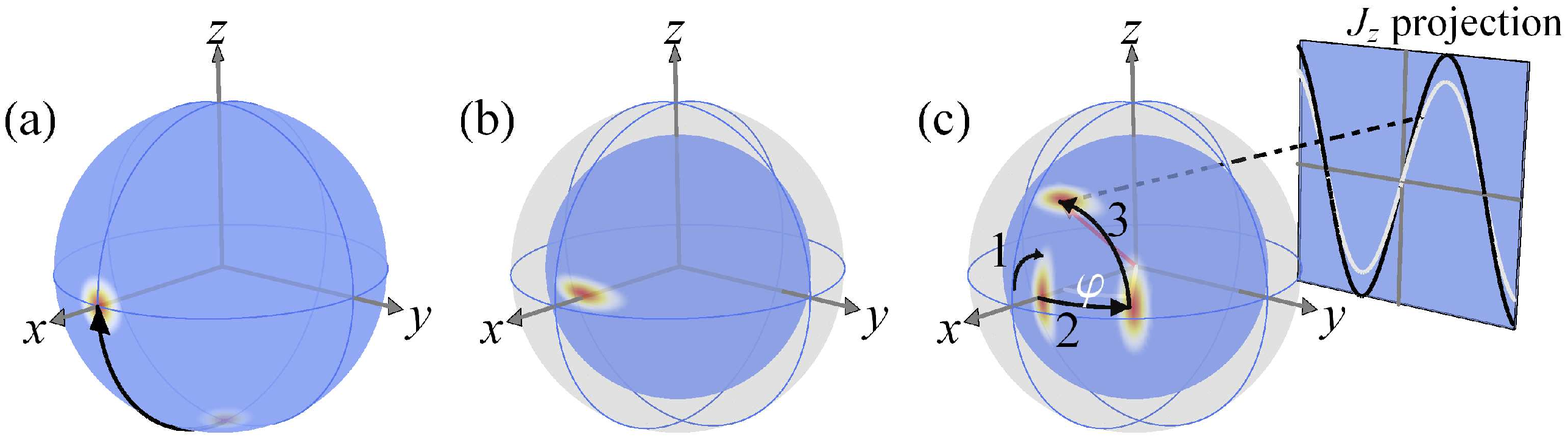}\\
  \includegraphics[width=10cm]{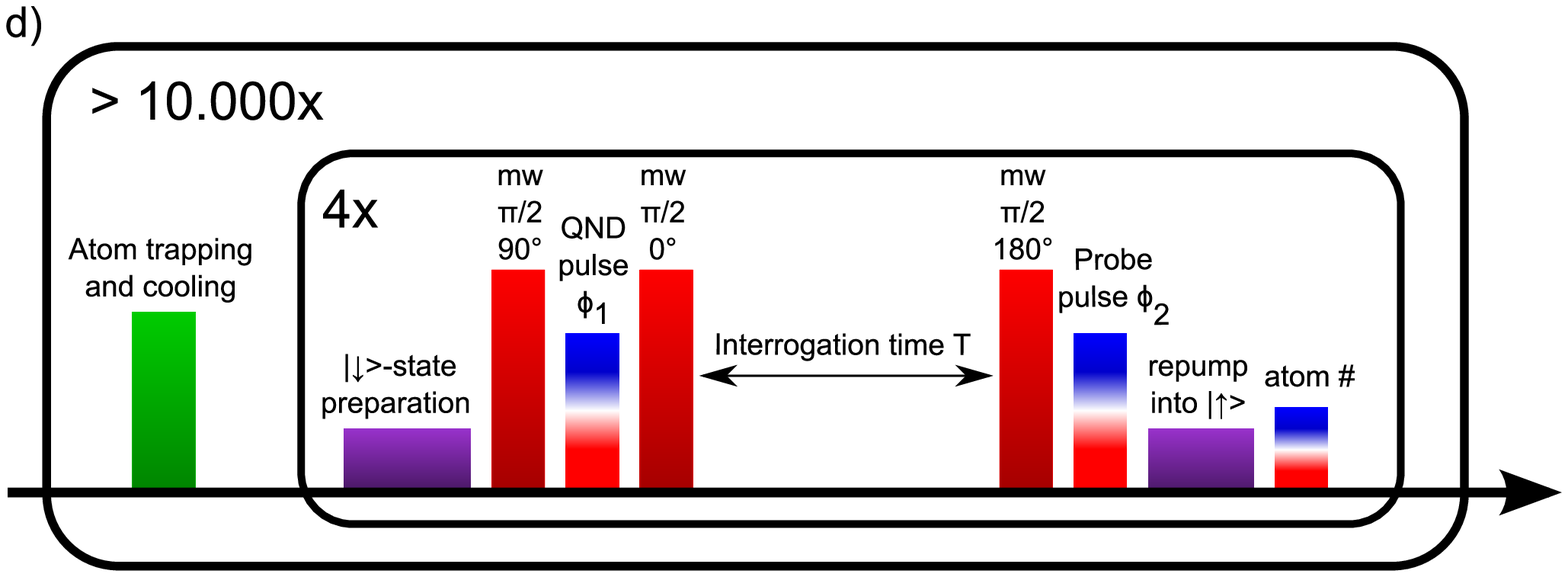}
  \caption{Ramsey sequence including squeezing in the Bloch sphere
    representation. The sequence starts with all atoms in
    $\ket{\downarrow}$. (a) A near resonant $\pi/2$-pulse is applied
    with phase $\vartheta_0=90^\circ$, placing the atoms in a coherent
    spin state along the $x$-axis. (b) The atomic state is squeezed
    along the $z$-direction by a weak QND measurement of the
    population difference.  The decoherence induced by the first QND
    measurement leads to a reduction of the Bloch sphere radius. (c.1)
    A second $\pi/2$-pulse with phase $\vartheta_1=0^\circ$ rotates
    the atomic state around the $x$-axis, thus converting the
    population squeezing into phase squeezing. (c.2) Free precession
    during interrogation time $T$, equivalent to a rotation around the
    $z$-axis by an angle $\varphi=\Delta\, T$. (c.3) A third
    $\pi/2$-pulse, with phase $\vartheta_2=180^\circ$, acts as a
    $\pi/2$-rotation around the $x$-axis. The final state has a
    $z$-axis component $\langle J_z \rangle=(1-\eta)\frac{N_A}{2}
    \sin\varphi$. The reduction of the atomic projection noise results
    in a reduced uncertainty in the measurement of $\phi_2$. (d)
    Diagram of the pulse sequence. }
  \label{fig:clockramseysq}
\end{figure}

We use the spin squeezing technique described in the previous section
to improve the precision of a Ramsey clock. The modified Ramsey
sequence is shown in Fig.~\ref{fig:clockramseysq}. As in a traditional
Ramsey sequence, all atoms are prepared in $\ket{\downarrow}$
initially. A near-resonant $\pi/2$-pulse with a phase of
$\vartheta_0=90^\circ$ and detuning $\Delta$ drives them into a CSS with macroscopic $J_x =
\frac{N_A}{2}$ (Fig.~\ref{fig:clockramseysq}.a). This state is then
squeezed along the $z$-direction by performing a weak QND measurement
of the pseudo-spin component $J_z$ (Fig.~\ref{fig:clockramseysq}.b).
This population-squeezed state is converted into a phase-squeezed
state by a second $\pi/2$-pulse with phase $\vartheta_1=0^\circ$ that
rotates the state around the $x$-axis. At this point, the Ramsey
interrogation time starts (Fig~\ref{fig:clockramseysq}.c.1).  We
let the atoms evolve freely for a time $T$, during which they acquire
a phase proportional to the microwave detuning $\varphi=\Delta\, T$
(Fig~\ref{fig:clockramseysq}.c.2). To stop the clock, a third
$\pi/2$-pulse with phase $180^\circ$ is applied, converting the
accumulated atomic phase shift $\varphi$ into a population difference
(Fig~\ref{fig:clockramseysq}.c.3). Finally, we measure the $J_z$-spin
component a second time and from the two measurement outcomes we
compute the conditional variance $\var{(\phi_2 - \zeta \phi_1)}$.

The first QND measurement of $J_z$ shown in
Fig.\ref{fig:clockramseysq}.b induces decoherence that reduces the
pseudo-spin vector length, and hence the Ramsey fringe height, so
that:
\begin{equation}
  \langle J_z\rangle=(1-\eta)\frac{N_A}{2}\sin\varphi,
  \label{eq:fringe}
\end{equation}
in contrast to the Ramsey fringe height in the absence of a QND
measurement: $\langle J_z\rangle = \frac{N_A}{2} \sin\varphi$.
Therefore, a QND measurement reduces the clock phase-sensitivity
(given by the Ramsey signal slope $d\phi/d\varphi)$ by a factor
$1-\eta$. Tailoring the QND measurement strength induces squeezing,
and thus improves the signal-to-projection-noise ratio by $1/\xi$ in
variance as given in Eq.~\ref{eq:wineland}.

We emphasize that atoms that have undergone spontaneous emissions into
the $m_F\neq 0$ states do not take part in the clock rotations: Due to
the magnetic bias field the microwave radiation only couples the
$\ket{F=3,m_F=0}$ to the $\ket{F=4,m_F=0}$ state. These atoms are
however part of the entangled state as they carry some part of the
information of the QND measurement outcome: Only when these atoms are
measured together with the $m_F=0$-atoms in the second measurement
there is no additional partition noise due to spontaneous emissions.
It is therefore important that in the absence of a phase shift
($\varphi=0$) the rotation operator corresponding to our microwave
clock pulse sequence commute with $J_z$, i.e. the population
difference $\Delta N$ is not changed.

\subsection{Low phase noise microwave source}

A projection noise limited atomic clock with $N_A=10^5$ atoms can
resolve atomic phase fluctuations as small as $\delta \varphi =
1/\sqrt{N_A} = \unit[3]{mrad}$. This poses strict requirements on the
phase noise of our microwave oscillator: during the interrogation time
its phase has to be much more stable than $1/\sqrt{N_A}$ so that our
clock performance is not limited by the oscillator.  For an oscillator
with a white phase noise spectrum and for a clock interrogation time
of $T=\unit[10]{\mu s}$ this translates into a required relative phase
noise power density lower than $\unit[-100]{dBc/Hz}$ over a
\unit[100]{kHz} sideband next to the \unit[9.192]{GHz} carrier.  This
is more than one order of magnitude smaller than the phase noise of
the Agilent HP8341B microwave synthesizer used in~\cite{appel2009}.

A key step in the implementation of the clock sequence therefore was
the construction of a low-noise synthesizer chain: a \unit[9]{GHz}
dielectric resonator oscillator (DRO-9.000-FR, Poseidon Scientific
Instruments) with a phase noise of
$\unit[-132]{dBc/Hz}@\unit[100]{kHz}$ is phase locked to an
oven-controlled \unit[500]{MHz} quartz oscillator~(OCXO) (MV87, Morion
Inc.) within a bandwidth of \unit[15]{kHz}. The OCXO itself is slowly
$(\approx\unit[10]{Hz})$ locked to a GPS reference. A direct digital
synthesis (DDS) board (Analog Devices AD9910) is clocked from the
frequency doubled OCXO and produces a \unit[192]{MHz} signal which is
mixed onto the DRO output; a microwave cavity resonator, with a
\unit[50]{MHz} width, filters out the upper \unit[9.192]{GHz} sideband.
By using the DDS to shape the microwave pulses we control the
microwave pulse duration with a precise timing in \unit[4]{ns} steps
and control the microwave phase digitally with a $ {\unit[2 \pi\cdot 2^{-16}]{rad}} $
resolution, allowing for complex and precise pulse sequences.

\subsection{Ramsey fringe decay}

\begin{figure}[ht] \centering
  \footnotesize (a)\includegraphics[width=5.5cm,angle=-90]{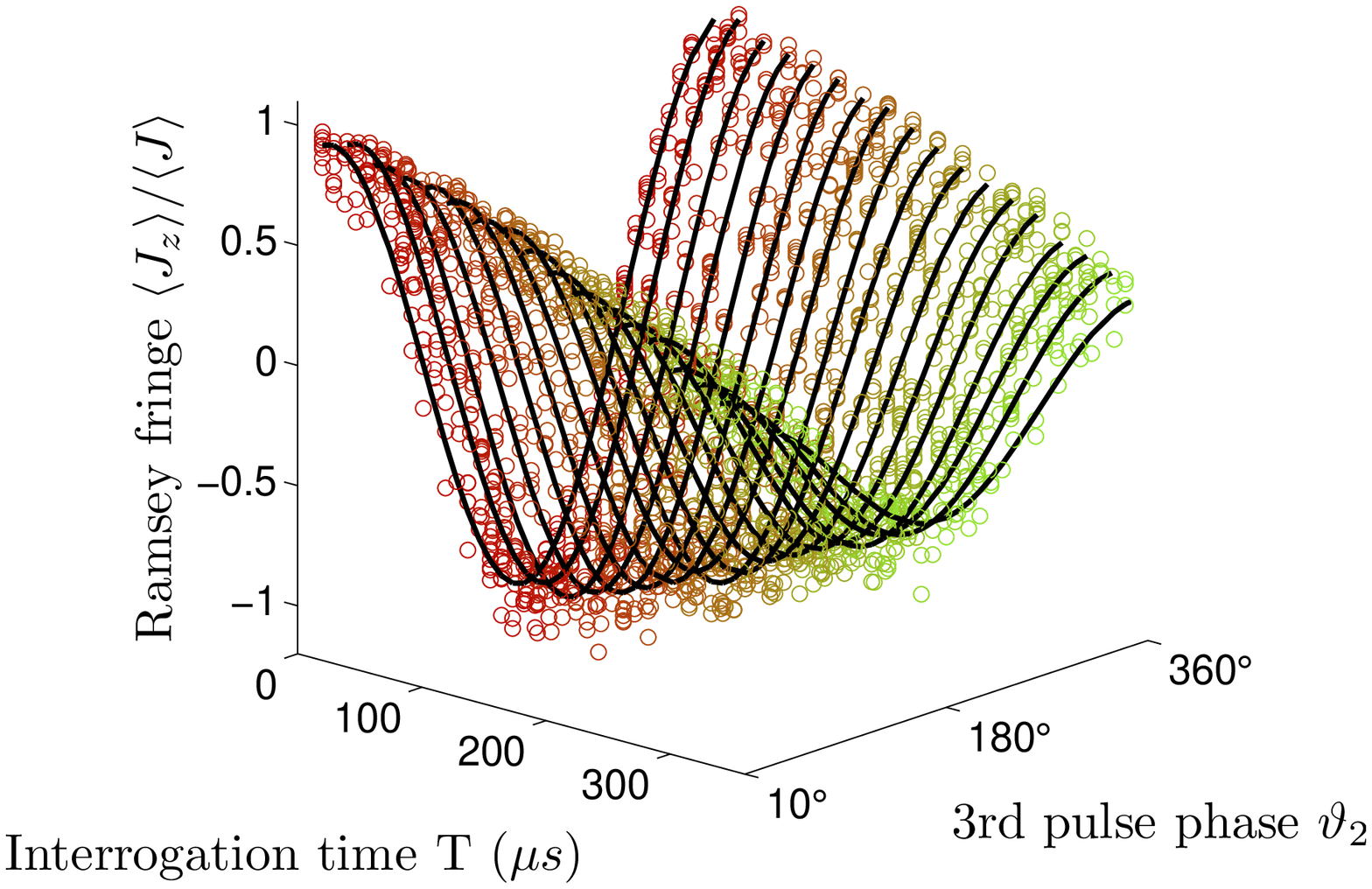}
  (b)\includegraphics[width=5cm,angle=-90]{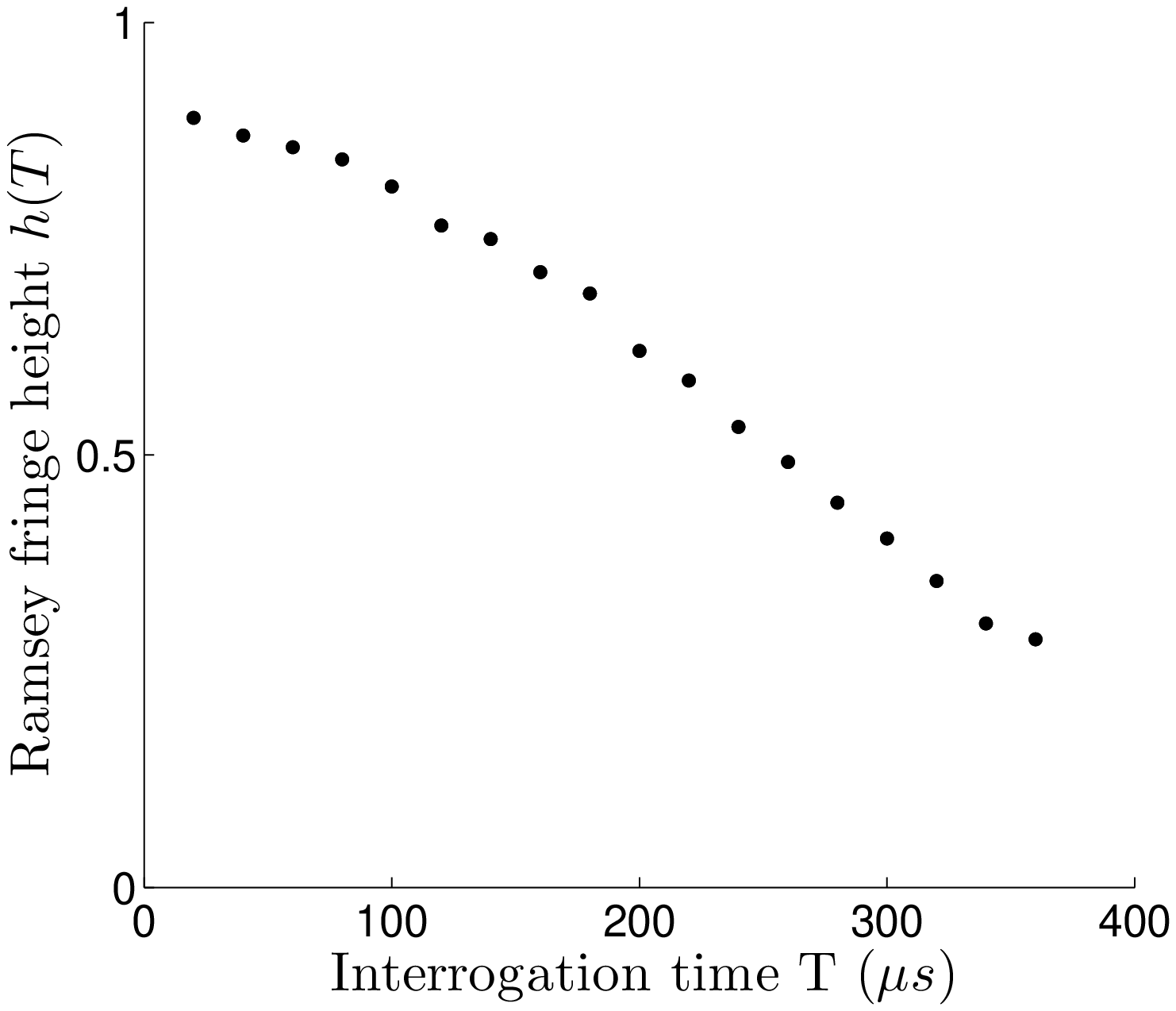}
  \caption{(a) Ramsey fringe for
    interrogation times ranging from $\unit[10]{\mu s}$ (red points)
    to $\unit[350]{\mu s}$ (green points). Solid lines: fits of type
    $\langle J_z\rangle /\langle J\rangle=h(T) \cos \vartheta_2$ to
    the data. (b) Fringe contrast $h(T)$ derived from the fits in (a).}
  \label{fig:fringedecay}
\end{figure}

We first operate our clock with the sequence described in
Fig.~\ref{fig:clockramseysq}, leaving out the first QND measurement.
The phase $\vartheta_2$ of the last microwave pulse is varied between
$0^\circ$ and $360^\circ$. The microwave frequency is set to resonance
(to within $10$~Hz), so we expect $\langle J_z \rangle=\frac{N_A}{2}
\cos \vartheta_2$.  In Fig.~\ref{fig:fringedecay}, we plot the
distribution of $J_z$-measurements normalized to $\langle J
\rangle=N_A/2$, for different interrogation times $T$ between
$\unit[10]{\mu s}$ and $\unit[350]{\mu s}$.  The differential ac Stark
shift induced by the Gaussian dipole trap potential causes spatially
inhomogeneous dephasing during the whole interrogation time $T$.  This
results in a decay of the Ramsey fringe contrast $h(T)$ (and therefore a decay of
the clock phase sensitivity) as $T$ increases, as shown in
Fig.~\ref{fig:fringedecay}.

\subsection{Ramsey sequence with squeezing}

\begin{figure}[ht] \centering
  \includegraphics[width=8cm,angle=-90]{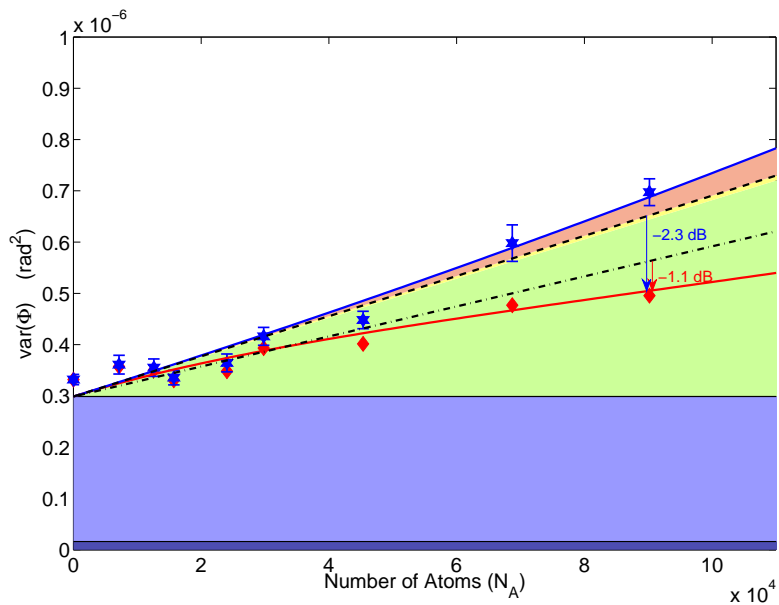}
  \caption{Noise contributions to the second measurement in the clock
    sequence, with an integration time of $T=10~\mu$s. The first
    measurement contains $7.1\cdot 10^6$ photons and induces a Bloch
    vector shortening of $13.5\%$.}
  \label{fig:clock_sq_10mus}
\end{figure}

We implement the full EAR sequence described in
Fig.~\ref{fig:clockramseysq}, and keep the last microwave pulse phase
$\vartheta_2=180^\circ$. The noise contributions to the second QND
measurement are shown in Fig.~\ref{fig:clock_sq_10mus}. With such a
short interrogation time ($10~\mu$s), only little classical noise is
added to the second measurement, compared to the first measurement
($\var (\phi_1)\approx\var (\phi_2)$). We attribute this classical
noise to microwave frequency noise and dipole trap intensity
fluctuations.

Using the first weak measurement $\phi_1$ to predict the outcome of
the second measurement $\phi_2$, we observe a metrologically
relevant noise reduction of $\xi=\unit[-1.1]{dB}$ for
$9\cdot10^4$~atoms. This gives us the signal-to-projection-noise ratio
improvement that we gained by running our clock with an entangled
atomic ensemble, compared to a standard clock operating with
unentangled atoms (see Fig.~\ref{fig:clock_sq_10mus}).  The
experimentally observed squeezing is lower than the expected value
$\xi_\text{lin}=[ (1-\eta)^2(1+\kappa^2) ]^{-1}=-2.2~$dB because of
the extra classical noise in the second measurement $\phi_2$.

Note that the obtained squeezing is different from the one shown in
Fig.~\ref{fig:squeezing}. Apart from the fact that the atom number
used in the clock experiment is lower, the reduction of the spin
squeezing can be explained by two experimental effects: firstly, the
atomic motion in the trap results in a decay of the correlations
$\cov(\phi_1,\phi_2)$ during the the longer time interval between the
two QND measurements $\phi_1,\phi_2$ in the clock sequence. Secondly,
in the Ramsey sequence, phase- and frequency fluctuations between
atoms and the microwave oscillator affect the $\phi_2$ measurement to
first order (cf. Eq.~\ref{eq:fringe}), whereas they only appear in
second order in the simple squeezing sequence shown in
Fig.\ref{fig:seqsqueezing}.

The first QND measurement $\phi_1$ also produces backaction
antisqueezing on the conjugate spin variable $J_x$. Classical
fluctuations in relative intensities of the two probe
colors lead to a fluctuating differential AC-Stark shift between the clock
levels.  This adds additional noise to $J_x$, so that the produced SSS
 is about \unit[10]{dB} more noisy in the (anti-squeezed) $J_x$
quadrature than a minimum uncertainty state. However, this excess
noise does not affect the clock precision since we only perform $J_z$
measurements.

\subsection{Frequency noise measurements}

The Ramsey sequence makes it possible to use an atomic ensemble either
as a clock, where the frequency of an oscillator is locked to the
transition frequency between the clock states
$\nu=(E_\uparrow-E_\downarrow)/h$, or as a sensor to measure
perturbations of the energy difference between these levels.  In our
experiment, the atomic ensemble is sensitive to fluctuations of the
frequency difference $\Delta$ between the cesium clock transition
and the microwave oscillator.

The atomic transition frequency of $\unit[9\, 192\, 630\, 500]{Hz}$ as
measured in our experiments has a systematic offset from the SI value
of $\unit[9\, 192\, 631\, 770]{Hz}$ mainly for two reasons: firstly,
the bias field of $\approx \unit[1]{G}$ causes a quadratic Zeeman
shift of $\unit[+427]{Hz/G^2}$ on the clock levels. Secondly, the
trapping light produces a differential ac~Stark shift of the order of
$\unit[-1700]{Hz}$ averaged over the atoms within the probe beam.
Noise in the phase evolution between the atoms and the microwave
oscillator leads to added classical noise in the $\phi_2$ measurement,
which scales as ${N_A}^2$.

The frequency noise components that an atomic clock can detect are
determined by the clock's cycle time $t_c$, and the interrogation time
$T$.  Uncorrelated frequency fluctuations from cycle to cycle result
in additional noise in the $\phi_2$-measurement that scales as $T^2$
in variance.  Phase noise between atoms and the microwave oscillator
during the interrogation time $T$, however, can follow a different
scaling behaviour.  Since we subtract the outcomes of successive MOT
loading-cycles, our clock is not sensitive to any frequency noise
slower than $t_c=\unit[5]{s}$.

\subsection{Influence of classical noise on the clock performance}

The quantum noise limited frequency sensitivity of an atomic clock
improves with increasing the interrogation time. However, achieving a
higher frequency sensitivity makes our experiment more susceptible to
classical noise. In order to evaluate the limitations of our
proof-of-principle experiment, we vary the interrogation time $T$ from
$\unit[10]{\mu s}$ to $\unit[310]{\mu s}$ by steps of $20~\mu$s, while
keeping the atom number approximately constant at $N_A=9\cdot10^4$.
From the optical phase shift measurements $\phi_1$ and $\phi_2$, we
can infer the atomic phase evolution by normalizing to the Ramsey
fringe amplitude and define:
\begin{equation}
  \label{eq:tilde}
  \tilde \varphi_2 = \frac{\phi_2}{A(T)} \qquad
  \tilde \varphi_{21} = \frac{\phi_2-\zeta \phi_1}{A(T)}.
\end{equation}
where $A(T) = \chi\, (1-\eta)\, h(T)\, N_A$ and $h(T)$ is the Ramsey
fringe contrast shown in Fig.~\ref{fig:fringedecay}.

\begin{figure}[ht] \centering
  \includegraphics[width=9cm]{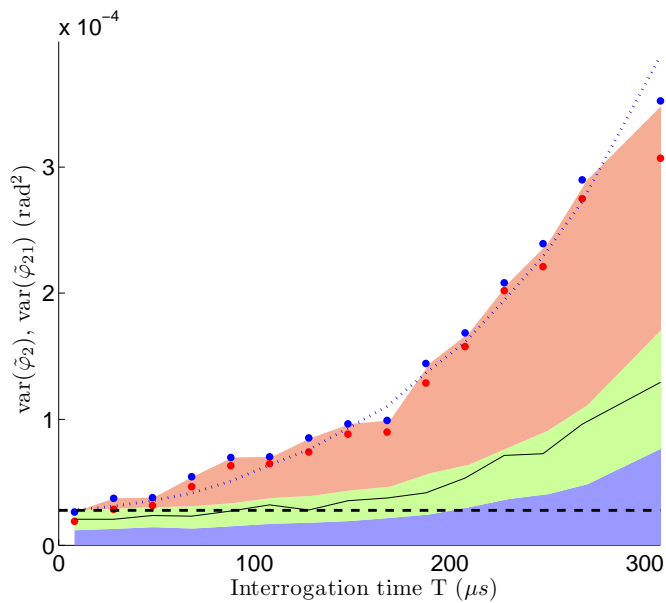}
  \caption{Contributions to the atomic phase noise as a function of
    the interrogation time $T$. Blue points: $\var(\tilde\varphi_2)$.
    Red points: $\var(\tilde\varphi_{21})$. For each interrogation
    time $T$, we determine the various noise contributions to
    $\var(\phi_2)$, as shown in Fig.~\ref{fig:clock_sq_10mus}.  Using
    Eq.~\ref{eq:tilde}, we obtain the noise contributions to
    $\var(\tilde\varphi_2)$. The different contributions to
    $\var(\tilde\varphi_2)$ are: the shot noise (blue area),
    projection noise (green area), and classical noise (red area).
    The dotted line is a quadratic fit to the classical noise
    contribution to $\var(\tilde\varphi_{21})$ (red area). The solid
    line shows the reduced projection noise which would be obtainable
    in the absence of classical noise. The dashed line demarcates the
    area in which the Wineland criterion (cf. Eq.~\ref{eq:wineland})
    is satisfied $\var(\tilde
    \varphi_{21})<1/N_A$, corresponding to the regime of
    metrologically relevant squeezing. In the absence of classical
    noise, the maximum interrogation time at which the Wineland
    criterion would be satisfied is given by the intersection between
    the dashed and solid lines at $T=\unit[90]{\mu s}$.}
  \label{fig:phasenoise}
\end{figure}

In Figure~\ref{fig:phasenoise}, we plot the measured atomic phase
noise variance $\var(\tilde \varphi_2)$ and the conditionally reduced
noise of the atomic phase $\var(\tilde \varphi_{21})$ as a function of
the interrogation time $T$. Only in the first data point
(corresponding to $T=\unit[10]{\mu s}$) we observe squeezing, i.e. the
measured noise in $\tilde \varphi_{21}$ is lower than that of a
traditionally operated atomic clock with an identical
signal-to-projection-noise ratio (i.e. with an atom number of
$N_A/\xi_\text{lin}$).

We observe a quadratic increase in the classical noise with $T$.  This
suggests that the detuning is roughly constant over the interrogation
time, but varies from cycle to cycle.  We attribute these fluctuations
to both intensity drifts in the dipole trap and variations of the
geometry of the atomic cloud. From the fit (solid green line) we infer
$\sqrt{\var(\Delta)} = \unit[7.5]{Hz}$ per cycle.

\section{Conclusion}

In summary, we have demonstrated the first entanglement-assisted
Ramsey clock, following the proposal of~\cite{appel2009}. We have
implemented a modified Ramsey sequence in which the spin state is
squeezed by means of a quantum non-demolition measurement. Squeezing
results in a metrologically relevant reduction of the noise variance by
$\unit[-1.1]{dB}$ with $9\cdot10^4$ atoms, compared to a traditional
atomic clock at the projection noise limit with the same number of
atoms. The present experimental developments have been made possible
by the introduction of a low phase noise microwave source.

The dual-color QND method implemented in this work is readily
applicable to optical clocks as well. The idea of using quantum
non-demolition measurements to generate entanglement and to improve
the precision of a clock has drawn the attention of the
ultra-precise-frequency-standards
community~\cite{meiser2008,lodewyck:Nondestructivemeasurement}.
Besides, non-destructive measurements have been shown to drastically
improve the duty cycle in a clock experiment, thereby reducing the
Dick effect~\cite{lodewyck:Nondestructivemeasurement}.

We have shown that for the present proof-of-principle experiment, the
improvement of the clock precision due to squeezing does not extend to
interrogation times beyond $\unit[10]{\mu s}$.  This is due to the
inhomogeneous broadening on the hyperfine transition induced by the
dipole trap, and the fact that large interrogation times make the
clock more sensitive to light shifts induced by cloud-geometry- and
intensity fluctuations in the dipole trap.  These issues could be
circumvented by turning to systems for which there exists a magic
wavelength, such as
strontium~\cite{katori03:Sr_magic_wavelength,ye08:_quant_state_ligt_insensitive_traps}
or ytterbium\cite{Barber08:Yb_magic_wavelength}.  For long
interrogation times, the clock performance also is limited by the
atomic motion, which can be counteracted by introducing a transverse
optical lattice.

The figure of merit for the QND-induced spin squeezing is the resonant
optical depth of the atomic ensemble. In the present experiment the
optical depth was limited to a value $\approx 10$. Implementation of QND-based
spin squeezing in state-of-the-art clocks will require solutions where
high optical depth can be combined with low collisional broadening,
such as, for example, optical lattices placed in a low finesse optical
cavity.

\section*{Acknowledgements}
The authors would like to thank J. H. M\"uller for fruitful and
inspiring discussions and Patrick Windpassinger and Ulrich Busk Hoff
for valuable contributions at the early stages of the experiment. This
research was supported by EU grants COMPAS, Q-ESSENCE, HIDEAS, and
QAP.

\section*{References}

\bibliography{clock4}

\begin{thebibliography}{10}%
\makeatletter
\providecommand \@ifxundefined [1]{%
 \ifx #1\undefined \expandafter \@firstoftwo
 \else \expandafter \@secondoftwo
\fi
}%
\providecommand \@ifnum [1]{%
 \ifnum #1\expandafter \@firstoftwo
 \else \expandafter \@secondoftwo
\fi
}%
\providecommand \enquote [1]{``#1''}%
\providecommand \bibnamefont  [1]{#1}%
\providecommand \bibfnamefont [1]{#1}%
\providecommand \citenamefont [1]{#1}%
\providecommand\href[0]{\@sanitize\@href}%
\providecommand\@href[1]{\endgroup\@@startlink{#1}\endgroup\@@href}%
\providecommand\@@href[1]{#1\@@endlink}%
\providecommand \@sanitize [0]{\begingroup\catcode`\&12\catcode`\#12\relax}%
\@ifxundefined \pdfoutput {\@firstoftwo}{%
 \@ifnum{\z@=\pdfoutput}{\@firstoftwo}{\@secondoftwo}%
}{%
 \providecommand\@@startlink[1]{\leavevmode\special{html:<a href="#1">}}%
 \providecommand\@@endlink[0]{\special{html:</a>}}%
}{%
 \providecommand\@@startlink[1]{%
  \leavevmode
  \pdfstartlink
   attr{/Border[0 0 1 ]/H/I/C[0 1 1]}%
   user{/Subtype/Link/A<</Type/Action/S/URI/URI(#1)>>}%
  \relax
 }%
 \providecommand\@@endlink[0]{\pdfendlink}%
}%
\providecommand \url  [0]{\begingroup\@sanitize \@url }%
\providecommand \@url [1]{\endgroup\@href {#1}{\urlprefix}}%
\providecommand \urlprefix [0]{URL }%
\providecommand \Eprint[0]{\href }%
\@ifxundefined \urlstyle {%
  \providecommand \doi [1]{doi:\discretionary{}{}{}#1}%
}{%
  \providecommand \doi [0]{doi:\discretionary{}{}{}\begingroup
  \urlstyle{rm}\Url }%
}%
\providecommand \doibase [0]{http://dx.doi.org/}%
\providecommand \Doi[1]{\href{\doibase#1}}%
\providecommand \selectlanguage [0]{\@gobble}%
\providecommand \bibinfo [0]{\@secondoftwo}%
\providecommand \bibfield [0]{\@secondoftwo}%
\providecommand \translation [1]{[#1]}%
\providecommand \BibitemOpen[0]{}%
\providecommand \bibitemStop [0]{}%
\providecommand \bibitemNoStop [0]{.\EOS\space}%
\providecommand \EOS [0]{\spacefactor3000\relax}%
\providecommand \BibitemShut [1]{\csname bibitem#1\endcsname}%
\bibitem{santarelli1999}%
  \BibitemOpen
  \bibfield{author}{%
  \bibinfo {author} {\bibfnamefont{G.}~\bibnamefont{Santarelli}}, \bibinfo
  {author} {\bibfnamefont{P.}~\bibnamefont{Laurent}}, \bibinfo {author}
  {\bibfnamefont{P.}~\bibnamefont{Lemonde}}, \bibinfo {author}
  {\bibfnamefont{A.}~\bibnamefont{Clairon}}, \bibinfo {author}
  {\bibfnamefont{A.~G.}\ \bibnamefont{Mann}}, \bibinfo {author}
  {\bibfnamefont{S.}~\bibnamefont{Chang}}, \bibinfo {author}
  {\bibfnamefont{A.~N.}\ \bibnamefont{Luiten}},\ and\ \bibinfo {author}
  {\bibfnamefont{C.}~\bibnamefont{Salomon}},\ }%
  \bibfield{journal}{%
  \Doi{10.1103/PhysRevLett.82.4619}{\bibinfo {journal} {Phys. Rev. Lett.}}\ }%
  \textbf{\bibinfo {volume} {82}},\ \bibinfo {pages} {4619} (\bibinfo {year}
  {1999})\BibitemShut{NoStop}%
\bibitem{wilpers2002}%
  \BibitemOpen
  \bibfield{author}{%
  \bibinfo {author} {\bibfnamefont{G.}~\bibnamefont{Wilpers}}, \bibinfo
  {author} {\bibfnamefont{T.}~\bibnamefont{Binnewies}}, \bibinfo {author}
  {\bibfnamefont{C.}~\bibnamefont{Degenhardt}}, \bibinfo {author}
  {\bibfnamefont{U.}~\bibnamefont{Sterr}}, \bibinfo {author}
  {\bibfnamefont{J.}~\bibnamefont{Helmcke}},\ and\ \bibinfo {author}
  {\bibfnamefont{F.}~\bibnamefont{Riehle}},\ }%
  \bibfield{journal}{%
  \Doi{10.1103/PhysRevLett.89.230801}{\bibinfo {journal} {Phys. Rev. Lett.}}\
  }%
  \textbf{\bibinfo {volume} {89}},\ \bibinfo {pages} {230801} (\bibinfo {year}
  {2002})\BibitemShut{NoStop}%
\bibitem{Ludlow:2008_SrLatticeClock}%
  \BibitemOpen
  \bibfield{author}{%
  \bibinfo {author} {\bibfnamefont{A.~D.}\ \bibnamefont{Ludlow}}, \bibinfo
  {author} {\bibfnamefont{T.}~\bibnamefont{Zelevinsky}}, \bibinfo {author}
  {\bibfnamefont{G.~K.}\ \bibnamefont{Campbell}}, \bibinfo {author}
  {\bibfnamefont{S.}~\bibnamefont{Blatt}}, \bibinfo {author}
  {\bibfnamefont{M.~M.}\ \bibnamefont{Boyd}}, \bibinfo {author}
  {\bibfnamefont{M.~H.~G.}\ \bibnamefont{de~Miranda}}, \bibinfo {author}
  {\bibfnamefont{M.~J.}\ \bibnamefont{Martin}}, \bibinfo {author}
  {\bibfnamefont{J.~W.}\ \bibnamefont{Thomsen}}, \bibinfo {author}
  {\bibfnamefont{S.~M.}\ \bibnamefont{Foreman}}, \bibinfo {author}
  {\bibfnamefont{J.}~\bibnamefont{Ye}}, \bibinfo {author}
  {\bibfnamefont{T.~M.}\ \bibnamefont{Fortier}}, \bibinfo {author}
  {\bibfnamefont{J.~E.}\ \bibnamefont{Stalnaker}}, \bibinfo {author}
  {\bibfnamefont{S.~A.}\ \bibnamefont{Diddams}}, \bibinfo {author}
  {\bibfnamefont{Y.~L.}\ \bibnamefont{Coq}}, \bibinfo {author}
  {\bibfnamefont{Z.~W.}\ \bibnamefont{Barber}}, \bibinfo {author}
  {\bibfnamefont{N.}~\bibnamefont{Poli}}, \bibinfo {author}
  {\bibfnamefont{N.~D.}\ \bibnamefont{Lemke}}, \bibinfo {author}
  {\bibfnamefont{K.~M.}\ \bibnamefont{Beck}},\ and\ \bibinfo {author}
  {\bibfnamefont{C.~W.}\ \bibnamefont{Oates}},\ }%
  \bibfield{journal}{%
  \Doi{10.1126/science.1153341}{\bibinfo {journal} {Science}}\ }%
  \textbf{\bibinfo {volume} {319}},\ \bibinfo {pages} {1805} (\bibinfo {year}
  {2008})\BibitemShut{NoStop}%
\bibitem{wineland1992}%
  \BibitemOpen
  \bibfield{author}{%
  \bibinfo {author} {\bibfnamefont{D.~J.}\ \bibnamefont{Wineland}}, \bibinfo
  {author} {\bibfnamefont{J.~J.}\ \bibnamefont{Bollinger}}, \bibinfo {author}
  {\bibfnamefont{W.~M.}\ \bibnamefont{Itano}}, \bibinfo {author}
  {\bibfnamefont{F.~L.}\ \bibnamefont{Moore}},\ and\ \bibinfo {author}
  {\bibfnamefont{D.~J.}\ \bibnamefont{Heinzen}},\ }%
  \bibfield{journal}{%
  \Doi{10.1103/PhysRevA.46.R6797}{\bibinfo {journal} {Phys. Rev. A}}\ }%
  \textbf{\bibinfo {volume} {46}},\ \bibinfo {pages} {R6797} (\bibinfo {year}
  {1992})\BibitemShut{NoStop}%
\bibitem{huelga1997}%
  \BibitemOpen
  \bibfield{author}{%
  \bibinfo {author} {\bibfnamefont{S.~F.}\ \bibnamefont{Huelga}}, \bibinfo
  {author} {\bibfnamefont{C.}~\bibnamefont{Macchiavello}}, \bibinfo {author}
  {\bibfnamefont{T.}~\bibnamefont{Pellizzari}}, \bibinfo {author}
  {\bibfnamefont{A.~K.}\ \bibnamefont{Ekert}}, \bibinfo {author}
  {\bibfnamefont{M.~B.}\ \bibnamefont{Plenio}},\ and\ \bibinfo {author}
  {\bibfnamefont{J.~I.}\ \bibnamefont{Cirac}},\ }%
  \bibfield{journal}{%
  \Doi{10.1103/PhysRevLett.79.3865}{\bibinfo {journal} {Phys. Rev. Lett.}}\ }%
  \textbf{\bibinfo {volume} {79}},\ \bibinfo {pages} {3865} (\bibinfo {year}
  {1997})\BibitemShut{NoStop}%
\bibitem{giovannetti04:_quant_enhan_measur}%
  \BibitemOpen
  \bibfield{author}{%
  \bibinfo {author} {\bibfnamefont{V.}~\bibnamefont{Giovannetti}}, \bibinfo
  {author} {\bibfnamefont{S.}~\bibnamefont{Lloyd}},\ and\ \bibinfo {author}
  {\bibfnamefont{L.}~\bibnamefont{Maccone}},\ }%
  \bibfield{journal}{%
  \Doi{10.1126/science.1104149}{\bibinfo {journal} {Science}}\ }%
  \textbf{\bibinfo {volume} {306}},\ \bibinfo {pages} {1330} (\bibinfo {year}
  {2004})\BibitemShut{NoStop}%
\bibitem{andre2004}%
  \BibitemOpen
  \bibfield{author}{%
  \bibinfo {author} {\bibfnamefont{A.}~\bibnamefont{Andr\'e}}, \bibinfo
  {author} {\bibfnamefont{A.~S.}\ \bibnamefont{S\o{}rensen}},\ and\ \bibinfo
  {author} {\bibfnamefont{M.~D.}\ \bibnamefont{Lukin}},\ }%
  \bibfield{journal}{%
  \Doi{10.1103/PhysRevLett.92.230801}{\bibinfo {journal} {Phys. Rev. Lett.}}\
  }%
  \textbf{\bibinfo {volume} {92}},\ \bibinfo {pages} {230801} (\bibinfo {year}
  {2004})\BibitemShut{NoStop}%
\bibitem{meiser2008}%
  \BibitemOpen
  \bibfield{author}{%
  \bibinfo {author} {\bibfnamefont{D.}~\bibnamefont{Meiser}}, \bibinfo {author}
  {\bibfnamefont{J.}~\bibnamefont{Ye}},\ and\ \bibinfo {author}
  {\bibfnamefont{M.~J.}\ \bibnamefont{Holland}},\ }%
  \bibfield{journal}{%
  \Doi{10.1088/1367-2630/10/7/073014}{\bibinfo {journal} {New J. Phys.}}\ }%
  \textbf{\bibinfo {volume} {10}},\ \bibinfo {pages} {073014} (\bibinfo {year}
  {2008})\BibitemShut{NoStop}%
\bibitem{meyer01:_ion_spin_squeezing}%
  \BibitemOpen
  \bibfield{author}{%
  \bibinfo {author} {\bibfnamefont{V.}~\bibnamefont{Meyer}}, \bibinfo {author}
  {\bibfnamefont{M.~A.}\ \bibnamefont{Rowe}}, \bibinfo {author}
  {\bibfnamefont{D.}~\bibnamefont{Kielpinski}}, \bibinfo {author}
  {\bibfnamefont{C.~A.}\ \bibnamefont{Sackett}}, \bibinfo {author}
  {\bibfnamefont{W.~M.}\ \bibnamefont{Itano}}, \bibinfo {author}
  {\bibfnamefont{C.}~\bibnamefont{Monroe}},\ and\ \bibinfo {author}
  {\bibfnamefont{D.~J.}\ \bibnamefont{Wineland}},\ }%
  \bibfield{journal}{%
  \Doi{10.1103/PhysRevLett.86.5870}{\bibinfo {journal} {Phys. Rev. Lett.}}\ }%
  \textbf{\bibinfo {volume} {86}},\ \bibinfo {pages} {5870} (\bibinfo {month}
  {Jun}\ \bibinfo {year} {2001})\BibitemShut{NoStop}%
\bibitem{wasilewski:_magnetometry}%
  \BibitemOpen
  \bibfield{author}{%
  \bibinfo {author} {\bibfnamefont{W.}~\bibnamefont{Wasilewski}}, \bibinfo
  {author} {\bibfnamefont{K.}~\bibnamefont{Jensen}}, \bibinfo {author}
  {\bibfnamefont{H.}~\bibnamefont{Krauter}}, \bibinfo {author}
  {\bibfnamefont{J.}~\bibnamefont{Renema}}, \bibinfo {author}
  {\bibfnamefont{M.~V.}\ \bibnamefont{Balabas}},\ and\ \bibinfo {author}
  {\bibfnamefont{E.}~\bibnamefont{Polzik}},\ }%
  \enquote{\bibinfo {title} {Quantum noise limited and entanglement-assisted
  magnetometry},}\
  \Eprint{http://arxiv.org/abs/0907.2453v3}{arXiv:0907.2453v3}\BibitemShut{NoS%
top}%
\bibitem{koschorreck-2009:_magnetometry}%
  \BibitemOpen
  \bibfield{author}{%
  \bibinfo {author} {\bibfnamefont{M.}~\bibnamefont{Koschorreck}}, \bibinfo
  {author} {\bibfnamefont{M.}~\bibnamefont{Napolitano}}, \bibinfo {author}
  {\bibfnamefont{B.}~\bibnamefont{Dubost}},\ and\ \bibinfo {author}
  {\bibfnamefont{M.~W.}\ \bibnamefont{Mitchell}},\ }%
  \enquote{\bibinfo {title} {Measurement of spin projection noise in broadband
  atomic magnetometry},}\
  \Eprint{http://arxiv.org/abs/0911.4491v1}{arXiv:0911.4491v1}\BibitemShut{NoS%
top}%
\bibitem{appel2009}%
  \BibitemOpen
  \bibfield{author}{%
  \bibinfo {author} {\bibfnamefont{J.}~\bibnamefont{Appel}}, \bibinfo {author}
  {\bibfnamefont{P.~J.}\ \bibnamefont{Windpassinger}}, \bibinfo {author}
  {\bibfnamefont{D.}~\bibnamefont{Oblak}}, \bibinfo {author}
  {\bibfnamefont{U.~B.}\ \bibnamefont{Hoff}},\ and\ \bibinfo {author}
  {\bibfnamefont{N.}~\bibnamefont{Kjærgaard}},\ }%
  \bibfield{journal}{%
  \Doi{10.1073/pnas.0901550106}{\bibinfo {journal} {Proc. Natl. Acad. Sci.}}\
  }%
  \textbf{\bibinfo {volume} {106}},\ \bibinfo {pages} {10960} (\bibinfo {year}
  {2009})\BibitemShut{NoStop}%
\bibitem{lodewyck:Nondestructivemeasurement}%
  \BibitemOpen
  \bibfield{author}{%
  \bibinfo {author} {\bibfnamefont{J.}~\bibnamefont{Lodewyck}}, \bibinfo
  {author} {\bibfnamefont{P.~G.}\ \bibnamefont{Westergaard}},\ and\ \bibinfo
  {author} {\bibfnamefont{P.}~\bibnamefont{Lemonde}},\ }%
  \bibfield{journal}{%
  \Doi{10.1103/PhysRevA.79.061401}{\bibinfo {journal} {Phys. Rev. A}}\ }%
  \textbf{\bibinfo {volume} {79}},\ \bibinfo {eid} {061401} (\bibinfo {year}
  {2009})\BibitemShut{NoStop}%
\bibitem{itano1993}%
  \BibitemOpen
  \bibfield{author}{%
  \bibinfo {author} {\bibfnamefont{W.~M.}\ \bibnamefont{Itano}}, \bibinfo
  {author} {\bibfnamefont{J.~C.}\ \bibnamefont{Bergquist}}, \bibinfo {author}
  {\bibfnamefont{J.~J.}\ \bibnamefont{Bollinger}}, \bibinfo {author}
  {\bibfnamefont{J.~M.}\ \bibnamefont{Gilligan}}, \bibinfo {author}
  {\bibfnamefont{D.~J.}\ \bibnamefont{Heinzen}}, \bibinfo {author}
  {\bibfnamefont{F.~L.}\ \bibnamefont{Moore}}, \bibinfo {author}
  {\bibfnamefont{M.~G.}\ \bibnamefont{Raizen}},\ and\ \bibinfo {author}
  {\bibfnamefont{D.~J.}\ \bibnamefont{Wineland}},\ }%
  \bibfield{journal}{%
  \Doi{10.1103/PhysRevA.47.3554}{\bibinfo {journal} {Phys. Rev. A}}\ }%
  \textbf{\bibinfo {volume} {47}},\ \bibinfo {pages} {3554} (\bibinfo {year}
  {1993})\BibitemShut{NoStop}%
\bibitem{sorensenduan2001:bec_entanglement}%
  \BibitemOpen
  \bibfield{author}{%
  \bibinfo {author} {\bibfnamefont{A.}~\bibnamefont{Sørensen}}, \bibinfo
  {author} {\bibfnamefont{L.-M.}\ \bibnamefont{Duan}}, \bibinfo {author}
  {\bibfnamefont{J.~I.}\ \bibnamefont{Cirac}},\ and\ \bibinfo {author}
  {\bibfnamefont{P.}~\bibnamefont{Zoller}},\ }%
  \bibfield{journal}{%
  \Doi{10.1038/35051038}{\bibinfo {journal} {Nature}}\ }%
  \textbf{\bibinfo {volume} {409}},\ \bibinfo {pages} {63} (\bibinfo {year}
  {2001})\BibitemShut{NoStop}%
\bibitem{esteve2008:bec_spinsqueezing}%
  \BibitemOpen
  \bibfield{author}{%
  \bibinfo {author} {\bibfnamefont{J.}~\bibnamefont{Est\`eve}}, \bibinfo
  {author} {\bibfnamefont{C.}~\bibnamefont{Gross}}, \bibinfo {author}
  {\bibfnamefont{A.}~\bibnamefont{Weller}}, \bibinfo {author}
  {\bibfnamefont{S.}~\bibnamefont{Giovanazzi}},\ and\ \bibinfo {author}
  {\bibfnamefont{M.~K.}\ \bibnamefont{Oberthaler}},\ }%
  \bibfield{journal}{%
  \Doi{10.1038/nature07332}{\bibinfo {journal} {Nature}}\ }%
  \textbf{\bibinfo {volume} {455}},\ \bibinfo {pages} {1216} (\bibinfo {year}
  {2008})\BibitemShut{NoStop}%
\bibitem{kuzmich97:_spin_squeez_ensem_atoms_illum_squeez_light}%
  \BibitemOpen
  \bibfield{author}{%
  \bibinfo {author} {\bibfnamefont{A.}~\bibnamefont{Kuzmich}}, \bibinfo
  {author} {\bibfnamefont{K.}~\bibnamefont{M\o{}lmer}},\ and\ \bibinfo {author}
  {\bibfnamefont{E.~S.}\ \bibnamefont{Polzik}},\ }%
  \bibfield{journal}{%
  \Doi{10.1103/PhysRevLett.79.4782}{\bibinfo {journal} {Phys. Rev. Lett.}}\ }%
  \textbf{\bibinfo {volume} {79}},\ \bibinfo {pages} {4782} (\bibinfo {year}
  {1997})\BibitemShut{NoStop}%
\bibitem{hald99:_spin_squeez_by_light}%
  \BibitemOpen
  \bibfield{author}{%
  \bibinfo {author} {\bibfnamefont{J.}~\bibnamefont{Hald}}, \bibinfo {author}
  {\bibfnamefont{J.~L.}\ \bibnamefont{S\o{}rensen}}, \bibinfo {author}
  {\bibfnamefont{C.}~\bibnamefont{Schori}},\ and\ \bibinfo {author}
  {\bibfnamefont{E.~S.}\ \bibnamefont{Polzik}},\ }%
  \bibfield{journal}{%
  \Doi{10.1103/PhysRevLett.83.1319}{\bibinfo {journal} {Phys. Rev. Lett.}}\ }%
  \textbf{\bibinfo {volume} {83}},\ \bibinfo {pages} {1319} (\bibinfo {year}
  {1999})\BibitemShut{NoStop}%
\bibitem{appel2008:quantum_memory_squeezed}%
  \BibitemOpen
  \bibfield{author}{%
  \bibinfo {author} {\bibfnamefont{J.}~\bibnamefont{Appel}}, \bibinfo {author}
  {\bibfnamefont{E.}~\bibnamefont{Figueroa}}, \bibinfo {author}
  {\bibfnamefont{D.}~\bibnamefont{Korystov}}, \bibinfo {author}
  {\bibfnamefont{M.}~\bibnamefont{Lobino}},\ and\ \bibinfo {author}
  {\bibfnamefont{A.~I.}\ \bibnamefont{Lvovsky}},\ }%
  \bibfield{journal}{%
  \Doi{10.1103/PhysRevLett.100.093602}{\bibinfo {journal} {Phys. Rev. Lett}}\
  }%
  \textbf{\bibinfo {volume} {100}},\ \bibinfo {eid} {093602} (\bibinfo {year}
  {2008})\BibitemShut{NoStop}%
\bibitem{honda:storage_squeezed}%
  \BibitemOpen
  \bibfield{author}{%
  \bibinfo {author} {\bibfnamefont{K.}~\bibnamefont{Honda}}, \bibinfo {author}
  {\bibfnamefont{D.}~\bibnamefont{Akamatsu}}, \bibinfo {author}
  {\bibfnamefont{M.}~\bibnamefont{Arikawa}}, \bibinfo {author}
  {\bibfnamefont{Y.}~\bibnamefont{Yokoi}}, \bibinfo {author}
  {\bibfnamefont{K.}~\bibnamefont{Akiba}}, \bibinfo {author}
  {\bibfnamefont{S.}~\bibnamefont{Nagatsuka}}, \bibinfo {author}
  {\bibfnamefont{T.}~\bibnamefont{Tanimura}}, \bibinfo {author}
  {\bibfnamefont{A.}~\bibnamefont{Furusawa}},\ and\ \bibinfo {author}
  {\bibfnamefont{M.}~\bibnamefont{Kozuma}},\ }%
  \bibfield{journal}{%
  \Doi{10.1103/PhysRevLett.100.093601}{\bibinfo {journal} {Phys. Rev. Lett.}}\
  }%
  \textbf{\bibinfo {volume} {100}},\ \bibinfo {eid} {093601} (\bibinfo {year}
  {2008})\BibitemShut{NoStop}%
\bibitem{grangier91:QND}%
  \BibitemOpen
  \bibfield{author}{%
  \bibinfo {author} {\bibfnamefont{P.}~\bibnamefont{Grangier}}, \bibinfo
  {author} {\bibfnamefont{J.-F.}\ \bibnamefont{Roch}},\ and\ \bibinfo {author}
  {\bibfnamefont{G.}~\bibnamefont{Roger}},\ }%
  \bibfield{journal}{%
  \Doi{10.1103/PhysRevLett.66.1418}{\bibinfo {journal} {Phys. Rev. Lett.}}\ }%
  \textbf{\bibinfo {volume} {66}},\ \bibinfo {pages} {1418} (\bibinfo {year}
  {1991})\BibitemShut{NoStop}%
\bibitem{kuzmich98:_atomicQND}%
  \BibitemOpen
  \bibfield{author}{%
  \bibinfo {author} {\bibfnamefont{A.}~\bibnamefont{Kuzmich}}, \bibinfo
  {author} {\bibfnamefont{N.~P.}\ \bibnamefont{Bigelow}},\ and\ \bibinfo
  {author} {\bibfnamefont{L.}~\bibnamefont{Mandel}},\ }%
  \bibfield{journal}{%
  \Doi{10.1209/epl/i1998-00277-9}{\bibinfo {journal} {Europhys. Lett.}}\ }%
  \textbf{\bibinfo {volume} {42}},\ \bibinfo {pages} {481} (\bibinfo {year}
  {1998})\BibitemShut{NoStop}%
\bibitem{kuzmich00:_spinsqueez_continuous}%
  \BibitemOpen
  \bibfield{author}{%
  \bibinfo {author} {\bibfnamefont{A.}~\bibnamefont{Kuzmich}}, \bibinfo
  {author} {\bibfnamefont{L.}~\bibnamefont{Mandel}},\ and\ \bibinfo {author}
  {\bibfnamefont{N.~P.}\ \bibnamefont{Bigelow}},\ }%
  \bibfield{journal}{%
  \Doi{10.1103/PhysRevLett.85.1594}{\bibinfo {journal} {Phys. Rev. Lett.}}\ }%
  \textbf{\bibinfo {volume} {85}},\ \bibinfo {pages} {1594} (\bibinfo {year}
  {2000})\BibitemShut{NoStop}%
\bibitem{chaudhury06:_contin_nondem_measur_cs_clock_trans_pseud}%
  \BibitemOpen
  \bibfield{author}{%
  \bibinfo {author} {\bibfnamefont{S.}~\bibnamefont{Chaudhury}}, \bibinfo
  {author} {\bibfnamefont{G.~A.}\ \bibnamefont{Smith}}, \bibinfo {author}
  {\bibfnamefont{K.}~\bibnamefont{Schulz}},\ and\ \bibinfo {author}
  {\bibfnamefont{P.~S.}\ \bibnamefont{Jessen}},\ }%
  \bibfield{journal}{%
  \Doi{10.1103/PhysRevLett.96.043001}{\bibinfo {journal} {Phys. Rev. Lett.}}\
  }%
  \textbf{\bibinfo {volume} {96}},\ \bibinfo {eid} {043001} (\bibinfo {year}
  {2006})\BibitemShut{NoStop}%
\bibitem{SchleierSmith2008}%
  \BibitemOpen
  \bibfield{author}{%
  \bibinfo {author} {\bibfnamefont{M.~H.}\ \bibnamefont{Schleier-Smith}},
  \bibinfo {author} {\bibfnamefont{I.~D.}\ \bibnamefont{Leroux}},\ and\
  \bibinfo {author} {\bibfnamefont{V.}~\bibnamefont{Vuleti{\'c}}},\ }%
  \bibfield{journal}{%
  \bibinfo {journal} {Phys. Rev. Lett.}\ }%
  \textbf{\bibinfo {volume} {104}},\ \bibinfo {pages} {073604} (\bibinfo {year}
  {2010})\BibitemShut{NoStop}%
\bibitem{takano09:_spin_squeez_cold_atomic_ensem}%
  \BibitemOpen
  \bibfield{author}{%
  \bibinfo {author} {\bibfnamefont{T.}~\bibnamefont{Takano}}, \bibinfo {author}
  {\bibfnamefont{M.}~\bibnamefont{Fuyama}}, \bibinfo {author}
  {\bibfnamefont{R.}~\bibnamefont{Namiki}},\ and\ \bibinfo {author}
  {\bibfnamefont{Y.}~\bibnamefont{Takahashi}},\ }%
  \bibfield{journal}{%
  \Doi{10.1103/PhysRevLett.102.033601}{\bibinfo {journal} {Phys. Rev. Lett.}}\
  }%
  \textbf{\bibinfo {volume} {102}},\ \bibinfo {eid} {033601} (\bibinfo {year}
  {2009})\BibitemShut{NoStop}%
\bibitem{julsgaard01:_entanglement}%
  \BibitemOpen
  \bibfield{author}{%
  \bibinfo {author} {\bibfnamefont{B.}~\bibnamefont{Julsgaard}}, \bibinfo
  {author} {\bibfnamefont{A.}~\bibnamefont{Kozhekin}},\ and\ \bibinfo {author}
  {\bibfnamefont{E.~S.}\ \bibnamefont{Polzik}},\ }%
  \bibfield{journal}{%
  \Doi{10.1038/35096524}{\bibinfo {journal} {Nature}}\ }%
  \textbf{\bibinfo {volume} {413}},\ \bibinfo {pages} {400} (\bibinfo {year}
  {2001})\BibitemShut{NoStop}%
\bibitem{saffman09:_spinsqueez_multicolor}%
  \BibitemOpen
  \bibfield{author}{%
  \bibinfo {author} {\bibfnamefont{M.}~\bibnamefont{Saffman}}, \bibinfo
  {author} {\bibfnamefont{D.}~\bibnamefont{Oblak}}, \bibinfo {author}
  {\bibfnamefont{J.}~\bibnamefont{Appel}},\ and\ \bibinfo {author}
  {\bibfnamefont{E.~S.}\ \bibnamefont{Polzik}},\ }%
  \bibfield{journal}{%
  \Doi{10.1103/PhysRevA.79.023831}{\bibinfo {journal} {Phys. Rev. A}}\ }%
  \textbf{\bibinfo {volume} {79}},\ \bibinfo {eid} {023831} (\bibinfo {year}
  {2009})\BibitemShut{NoStop}%
\bibitem{windpassinger08:_nondes_probin_rabi_oscil_cesium}%
  \BibitemOpen
  \bibfield{author}{%
  \bibinfo {author} {\bibfnamefont{P.~J.}\ \bibnamefont{Windpassinger}},
  \bibinfo {author} {\bibfnamefont{D.}~\bibnamefont{Oblak}}, \bibinfo {author}
  {\bibfnamefont{P.~G.}\ \bibnamefont{Petrov}}, \bibinfo {author}
  {\bibfnamefont{M.}~\bibnamefont{Kubasik}}, \bibinfo {author}
  {\bibfnamefont{M.}~\bibnamefont{Saffman}}, \bibinfo {author}
  {\bibfnamefont{C.~L.~G.}\ \bibnamefont{Alzar}}, \bibinfo {author}
  {\bibfnamefont{J.}~\bibnamefont{Appel}}, \bibinfo {author}
  {\bibfnamefont{J.~H.}\ \bibnamefont{M\"{u}ller}}, \bibinfo {author}
  {\bibfnamefont{N.}~\bibnamefont{Kjærgaard}},\ and\ \bibinfo {author}
  {\bibfnamefont{E.~S.}\ \bibnamefont{Polzik}},\ }%
  \bibfield{journal}{%
  \Doi{10.1103/PhysRevLett.100.103601}{\bibinfo {journal} {Phys. Rev. Lett.}}\
  }%
  \textbf{\bibinfo {volume} {100}},\ \bibinfo {eid} {103601} (\bibinfo {year}
  {2008})\BibitemShut{NoStop}%
\bibitem{Bibnote1}%
  \BibitemOpen
  \bibinfo {note} {In \cite {appel2009} we used opposite input ports for the
  two probe colors and operated the interferometer in its white-light position
  $\Delta L=0$ to minimize sensitivity to differential probe frequency noise.
  The method described here allows us to feed light of both probe colors into
  the interferometer through one common single mode fiber. This eliminates a
  possible spatial mode mismatch, which leads to spatially inhomogeneous
  differential AC-Stark shifts across the atomic sample. The differential probe
  frequency is controlled tightly using an optical phase lock \cite
  {appel09:_pll}.}\BibitemShut{Stop}%
\bibitem{appel09:_pll}%
  \BibitemOpen
  \bibfield{author}{%
  \bibinfo {author} {\bibfnamefont{J.}~\bibnamefont{Appel}}, \bibinfo {author}
  {\bibfnamefont{A.}~\bibnamefont{MacRae}},\ and\ \bibinfo {author}
  {\bibfnamefont{A.~I.}\ \bibnamefont{Lvovsky}},\ }%
  \bibfield{journal}{%
  \Doi{10.1088/0957-0233/20/5/055302}{\bibinfo {journal} {Meas. Sci.
  Technol.}}\ }%
  \textbf{\bibinfo {volume} {20}},\ \bibinfo {pages} {055302} (\bibinfo {year}
  {2009})\BibitemShut{NoStop}%
\bibitem{windpassinger09:_squeez}%
  \BibitemOpen
  \bibfield{author}{%
  \bibinfo {author} {\bibfnamefont{P.~J.}\ \bibnamefont{Windpassinger}},
  \bibinfo {author} {\bibfnamefont{D.}~\bibnamefont{Oblak}}, \bibinfo {author}
  {\bibfnamefont{U.~B.}\ \bibnamefont{Hoff}}, \bibinfo {author}
  {\bibfnamefont{A.}~\bibnamefont{Louchet}}, \bibinfo {author}
  {\bibfnamefont{J.}~\bibnamefont{Appel}}, \bibinfo {author}
  {\bibfnamefont{N.}~\bibnamefont{Kjærgaard}},\ and\ \bibinfo {author}
  {\bibfnamefont{E.~S.}\ \bibnamefont{Polzik}},\ }%
  \bibfield{journal}{%
  \Doi{10.1080/09500340903033682}{\bibinfo {journal} {J. Mod. Opt.}}\ }%
  \textbf{\bibinfo {volume} {56}},\ \bibinfo {pages} {1993} (\bibinfo {year}
  {2009})\BibitemShut{NoStop}%
\bibitem{oblak08:_echo_gauss}%
  \BibitemOpen
  \bibfield{author}{%
  \bibinfo {author} {\bibfnamefont{D.}~\bibnamefont{Oblak}}, \bibinfo {author}
  {\bibfnamefont{J.}~\bibnamefont{Appel}}, \bibinfo {author}
  {\bibfnamefont{P.}~\bibnamefont{Windpassinger}}, \bibinfo {author}
  {\bibfnamefont{U.}~\bibnamefont{Hoff}}, \bibinfo {author}
  {\bibfnamefont{N.}~\bibnamefont{Kjærgaard}},\ and\ \bibinfo {author}
  {\bibfnamefont{E.}~\bibnamefont{Polzik}},\ }%
  \bibfield{journal}{%
  \Doi{10.1140/epjd/e2008-00192-1}{\bibinfo {journal} {Eur. Phys. J. D}}\ }%
  \textbf{\bibinfo {volume} {50}},\ \bibinfo {pages} {67} (\bibinfo {year}
  {2008})\BibitemShut{NoStop}%
\bibitem{katori03:Sr_magic_wavelength}%
  \BibitemOpen
  \bibfield{author}{%
  \bibinfo {author} {\bibfnamefont{H.}~\bibnamefont{Katori}}, \bibinfo {author}
  {\bibfnamefont{M.}~\bibnamefont{Takamoto}}, \bibinfo {author}
  {\bibfnamefont{V.~G.}\ \bibnamefont{Pal\char39{}chikov}},\ and\ \bibinfo
  {author} {\bibfnamefont{V.~D.}\ \bibnamefont{Ovsiannikov}},\ }%
  \bibfield{journal}{%
  \Doi{10.1103/PhysRevLett.91.173005}{\bibinfo {journal} {Phys. Rev. Lett.}}\
  }%
  \textbf{\bibinfo {volume} {91}},\ \bibinfo {pages} {173005} (\bibinfo {year}
  {2003})\BibitemShut{NoStop}%
\bibitem{ye08:_quant_state_ligt_insensitive_traps}%
  \BibitemOpen
  \bibfield{author}{%
  \bibinfo {author} {\bibfnamefont{J.}~\bibnamefont{Ye}}, \bibinfo {author}
  {\bibfnamefont{H.~J.}\ \bibnamefont{Kimble}},\ and\ \bibinfo {author}
  {\bibfnamefont{H.}~\bibnamefont{Katori}},\ }%
  \bibfield{journal}{%
  \Doi{10.1126/science.1148259}{\bibinfo {journal} {Science}}\ }%
  \textbf{\bibinfo {volume} {320}},\ \bibinfo {pages} {1734} (\bibinfo {year}
  {2008})\BibitemShut{NoStop}%
\bibitem{Barber08:Yb_magic_wavelength}%
  \BibitemOpen
  \bibfield{author}{%
  \bibinfo {author} {\bibfnamefont{Z.~W.}\ \bibnamefont{Barber}}, \bibinfo
  {author} {\bibfnamefont{J.~E.}\ \bibnamefont{Stalnaker}}, \bibinfo {author}
  {\bibfnamefont{N.~D.}\ \bibnamefont{Lemke}}, \bibinfo {author}
  {\bibfnamefont{N.}~\bibnamefont{Poli}}, \bibinfo {author}
  {\bibfnamefont{C.~W.}\ \bibnamefont{Oates}}, \bibinfo {author}
  {\bibfnamefont{T.~M.}\ \bibnamefont{Fortier}}, \bibinfo {author}
  {\bibfnamefont{S.~A.}\ \bibnamefont{Diddams}}, \bibinfo {author}
  {\bibfnamefont{L.}~\bibnamefont{Hollberg}}, \bibinfo {author}
  {\bibfnamefont{C.~W.}\ \bibnamefont{Hoyt}}, \bibinfo {author}
  {\bibfnamefont{A.~V.}\ \bibnamefont{Taichenachev}},\ and\ \bibinfo {author}
  {\bibfnamefont{V.~I.}\ \bibnamefont{Yudin}},\ }%
  \bibfield{journal}{%
  \Doi{10.1103/PhysRevLett.100.103002}{\bibinfo {journal} {Phys. Rev. Lett.}}\
  }%
  \textbf{\bibinfo {volume} {100}},\ \bibinfo {pages} {103002} (\bibinfo
  {month} {Mar}\ \bibinfo {year} {2008})\BibitemShut{NoStop}%
\end{thebibliography}%

\end{document}